\documentclass[aps,prl,twocolumn,
superscriptaddress,nofootinbib,longbibliography]{revtex4-2}

\usepackage[utf8]{inputenc} 
\usepackage{amsmath,amssymb,bm}
\usepackage{mathtools}
\usepackage{graphicx,subfig,multirow,xcolor}
\usepackage[colorlinks=true,
            linkcolor=blue,citecolor=blue,urlcolor=blue]{hyperref}
\usepackage{placeins}
\usepackage{tikz}
\usepackage{tikz-3dplot}

\newcommand{\diff}{\mathrm d}
\newcommand{\rp}{r_{\parallel}}
\newcommand{\rpp}{\bm r_{\perp}}
\newcommand{\rppmag}{r_{\perp}}
\DeclareMathOperator{\arcosh}{arcosh}

\newcommand{\sUV}{\epsilon}             

\newcommand{\C}{\mathcal{C}}

\def\del{{\partial}}

\newcommand{\be}{\begin{equation}}
\newcommand{\ee}{\end{equation}}
\newcommand{\bea}{\begin{eqnarray}}
\newcommand{\eea}{\end{eqnarray}}

\begin{document}

\title{Entanglement, trace anomaly, and confinement in QCD}

\author{Kiminad A. Mamo}
\thanks{Contact author: \href{mailto:kamamo@wm.edu}{kamamo@wm.edu,}
\href{mailto:kamamo@jlab.org}{kamamo@jlab.org}}
\affiliation{Department of Physics, William and Mary, Williamsburg VA 23187, USA}
\affiliation{Theory Center, Jefferson Lab, Newport News, VA 23606, USA}

\date{December 5, 2025}

\begin{abstract}
We formulate \emph{confinement} in quantum chromodynamics (QCD) as an \hbox{entropic} surface phenomenon. Quark and gluon quantum information is localized on a transverse, \emph{entangling two-sphere} of radius \(R_{\!EE}\); at this radius the QCD vacuum---partitioned by a hadron into interior and exterior regions---reaches its maximal entanglement entropy. Lattice-QCD determinations of the scalar (trace) gravitational form factors fix both \(R_{\!EE}\) and the transverse trace-anomaly density \(\rho_h(R_{\!EE})\), yielding a \emph{parameter-free} slope \(c_h=8\pi^2 R_{\!EE}^2 \rho_h(R_{\!EE})\) and a mechanical entropy \(S_{\!EE}(y)=c_h\,y\) that grows linearly with rapidity \(y\). The entropy gradient \(\partial_{R} S_{\!EE}\) changes sign at \(R_{\!EE}\): it pushes colored degrees of freedom outward for \(r<R_{\!EE}\) and pulls them inward for \(r>R_{\!EE}\), thereby localizing them on the codimension-2 entangling two-sphere \(\Sigma_\perp=S^{2}_{R_{\!EE}}\) (which, in the infinite-momentum frame (IMF), projects onto the transverse plane)—the “information wall.” This provides a high-energy (large-\(y\)) entropic confinement diagnostic that complements—rather than replaces—Wilson’s area-law criterion, which probes long-distance dynamics near the rest frame (\(y\!\to\!0\)). Imposing unitarity on an entropic ansatz for the amplitude yields \(\sigma(s)\!\propto\!y^{\delta}\). World data \emph{favor} \(\delta=2\) for elastic \(pp\,(p\bar p)\) scattering and heavy-quark photoproduction, whereas \(\phi\) photoproduction favors a softer \(\delta=0.387\).
All extracted cross sections remain well below the Froissart--Martin bound.
These results provide a confinement criterion \emph{quantified} directly from nonperturbative QCD inputs, unifying the trace anomaly, entanglement entropy, and high-energy scattering within a single quantitative framework.
\end{abstract}

\maketitle

\textit{Introduction.—}
Why do colored quarks and gluons never emerge from hadrons?
The textbook answer invokes an \emph{area law} for large Wilson loops,
which diagnoses a linear confining potential in the infrared (IR) \cite{Wilson:1974sk}. Yet that
criterion carries no information about ultraviolet (UV) dynamics, where
asymptotic freedom reigns.

Here we propose a complementary, UV–sensitive diagnostic: \emph{entropic confinement}. Quantum information carried by quarks and gluons localizes on a codimension-2 entangling surface of fixed radius \(R_{\!EE}\). Geometrically, the entangling surface is the two–sphere \(\Sigma_\perp=S^2_R\) equipped with the induced metric \(h_{ab}\) and area element \(\sqrt{h}\,d^2x\). The universal contribution to the von Neumann entropy of the reduced state is then controlled by the quantum chromodynamics (QCD) trace anomaly on \(\Sigma_\perp\) \cite{Rosenhaus:2014nha,Ben-Ami:2015zsa}:
\bea
S_{\!EE}^{\rm univ}(R)&=&2\pi\ln\!\frac{R}{\epsilon}\!
   \int_{\Sigma_{\perp}}\!\sqrt{h}\,
   \langle T^{\mu}{}_{\mu}\rangle\nonumber\\
   &=&2\pi\,
     \ln\!\frac{R}{\sUV}
     \int_{r=R}\!R^2\,\diff\Omega\;
        \langle T^{\mu}{}_{\mu}\rangle,
\label{eq:Weyl_intro}
\eea
as established by general arguments in relativistic QFT~\cite{Bombelli:1986rw,Srednicki:1993im,Kabat:1994vj,Casini:2006es,Ryu:2006bv,Ryu:2006ef,Klebanov:2007ws,Solodukhin:2008dh,Casini:2009sr,Casini:2011kv,Rosenhaus:2014woa}.
See the Supplemental Material (included at the end of this arXiv version) for a detailed derivation of Eq.~\eqref{eq:Weyl_intro}.
In the infinite-momentum frame (IMF), the two-sphere \(S^2_R\) projects onto the transverse (impact-parameter) plane, and the surface measure reduces to the standard light-front impact-parameter measure used to define parton densities~\cite{Soper:1976jc,Burkardt:2000za}.
We use this mapping solely to express densities; the underlying entangling surface remains \(S^2_R\).

By “mechanical entropy’’ we mean the \emph{universal logarithmic} term \(S_{\!EE}^{\rm univ}\) evaluated under two physical renormalization conditions: (i) a maximal-entropy radius \(R=R_{\!EE}\) set by the hadron; and (ii) a Lorentz-contracted cutoff \(\epsilon=R_{\!EE}e^{-y}\) for boost rapidity \(y\). With these choices, spherical symmetry on \(\Sigma_\perp\) gives
\(\int_{\Sigma_\perp}\!\sqrt{h}\,\langle T^{\mu}{}_{\mu}\rangle
=4\pi R_{\!EE}^2\,\rho_h(R_{\!EE})\),
so that
\be
S_{\!EE}(y)=c_h\,y,
\qquad
c_h=8\pi^{2}R_{\!EE}^{2}\rho_h(R_{\!EE}).
\label{eq:SEE_boost}
\ee
At \(y=0\) the \emph{universal logarithmic} contribution is tuned to vanish by construction since $\epsilon=R_{\!EE}$ (therefore, our renormalization choice is valid only for large-$y$); this does not imply that the full state is pure.

For complementary 1+1D derivations of the linear-in-$y$ growth of entanglement, see \cite{Stoffers:2012mn,Liu:2018gae,Kharzeev:2017qzs,Gursoy:2023hge}. In particular, Ref.~\cite{Gursoy:2023hge} computes the energy-momentum tensor (EMT) of a “hadron-like’’ primary of dimension $h$ and finds (their Eq.~3.15) in pure 1+1D CFT
\(T_{++}(x_+)=C_h/x_+^{2}\) with \(C_h=h\); in the mass-deformed case (their Eq.~3.19) one reads off the same coefficient from \(C_h=x_{+}^{2}T_{++}(x_+)\) at $x_+\!\to\!0,\infty$. This gives a universal slope \(S_{EE}^{\rm univ}(y)=2\pi\,C_h\,y\), implying the dictionary \(C_h=c_h/(2\pi)\) when compared to Eq.~\eqref{eq:SEE_boost}. Our result is the 3+1D QCD realization: we determine \(c_h\) \emph{nonperturbatively} from the hadron’s trace anomaly localized on the entangling two-sphere, \(c_h=8\pi^{2}R_{EE}^{2}\rho_h(R_{EE})\), using lattice-QCD gravitational form factor (GFF) inputs in 2D impact-parameter space Eq.~\eqref{anomlyDensity}.

\begin{figure}[t]
  \centering
  \includegraphics[width=0.80\linewidth]{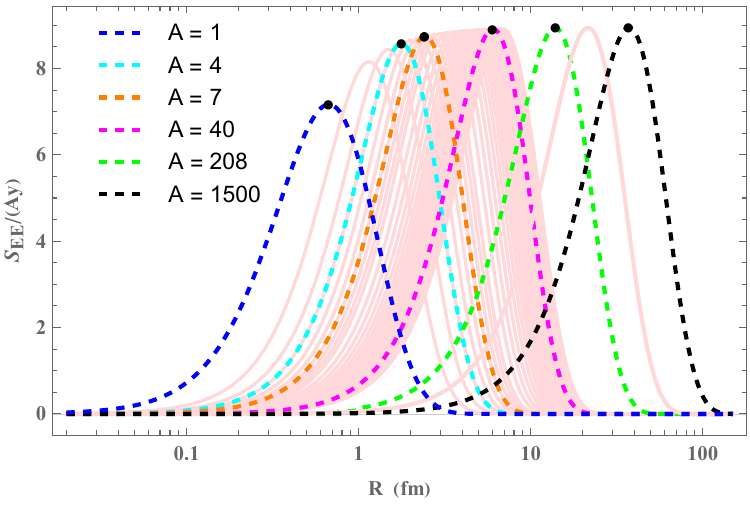}
  \caption{Mechanical entanglement entropy per nucleon,
  \(S_{\!EE}/(A\,y)\), versus entangling radius \(R\).
  The light band spans \(2\!\le\!A\!\le\!500\); dashed curves highlight
  representative nuclei.  All curves level off at
  \(S_{\!EE}/(A\,y)\!\simeq\!8.94\), signaling rapid saturation with
  \(A\).}
  \label{fig:SEE_nuclei}
\end{figure}

It is convenient to \emph{parametrize} the slope \(c_h\) in terms of an \emph{operational} (nonperturbative) occupation number on the information wall:
\begin{equation}
\frac{c_h}{8\pi^2 R_{\!EE}^2\,}\ \stackrel{\text{op.}}{=}\
\int_0^1\!dx\int d^2k_\perp\int d^2b_\perp\,
\delta_\Sigma(b_\perp)\,n_{q+g}(x,\bm{k}_\perp,\bm{b}_\perp)\,,
\label{eq:occupational}
\end{equation}
where $\delta_\Sigma(b_\perp)\equiv\frac{\delta(b_\perp-R_{\!EE})}{2\pi R_{\!EE}}$, and \(n_{q+g}(x,\bm{k_\perp},\bm{b}_\perp)\) denotes the total (sea-quark\,+\,gluon) occupation number with longitudinal momentum fraction \(x\) and transverse momentum \(\bm{k_\perp}\) per unit transverse phase-space area \(d^2b_\perp\); this identification serves as a nonperturbative analog of the gluon occupation number in the CGC/BFKL framework~\cite{Kuraev:1977fs,Balitsky:1978ic,Lipatov:1996ts,Jalilian-Marian:1997qno,Jalilian-Marian:1997jhx,Jalilian-Marian:1997ubg,Kovner:1999bj,Kovner:2000pt,Weigert:2000gi,Iancu:2000hn,Iancu:2001ad,Ferreiro:2001qy,Balitsky:1995ub,Balitsky:1998kc,Mueller:1993rr,McLerran:1993ni,McLerran:1993ka,Gelis:2010nm}.
Also note that \(c_h\) per nucleon shows rapid saturation with increasing mass number \(A\) in nuclei; see Fig.~\ref{fig:SEE_nuclei}.

Formally, we define the hadron’s entanglement by starting from the vacuum density operator \(|0\rangle\!\langle0|\) and \emph{tracing out} all quark- and gluon-field modes whose support lies \emph{outside} a sphere of radius \(R\), see Fig.~\ref{fig:A-B-partition} in the Supplemental Material.
The reduced density matrix admits the Euclidean path-integral representation~\cite{Rosenhaus:2014nha}
\be\label{redpath}
\hat\rho_{<}(R)\;=\;\mathrm{Tr}_{r>R}\,
  |0\rangle\!\langle0| = {1\over \mathcal{N}}\int_{\substack{\phi(\C_>)=\phi_>\\ \phi(\C_<)=\phi_<}} \mathcal{D}\phi \, e^{-I(\phi)}\,,
\ee
which describes the quantum state of the interior region (\(r<R\)). Here \(\C_{<\,\text{or}\,>}\) denote the two sides of a three-dimensional cut \(\mathcal{C}\) with boundary \(\del\,\mathcal{C}=\Sigma_{\perp}\); \(\phi\) collectively denotes all fields of the quantum field theory (QFT), and \(\phi_{<\,\text{or}\,>}\) are fixed field configurations on \(\C_{<\,\text{or}\,>}\). The eigenvalues \(\{\lambda_{n}(R)\}\) of \(\hat\rho_{<}(R)\) form the \emph{entanglement spectrum}. The von Neumann entropy \(S_{\!EE}(R)=-\sum_{n}\lambda_{n}\ln\lambda_{n}\) reduces, in the continuum limit, to Eq.~\eqref{eq:SEE_boost}, with \emph{spectral density} \(\rho_{h}(b_\perp)=\langle T^{\mu}{}_{\mu}(b_\perp)\rangle\) that appears in the QCD trace anomaly density in impact-parameter space Eq.~\eqref{anomlyDensity}.

Percent-level lattice QCD calculations of the \emph{scalar} (trace)
gravitational form factors (GFFs)
\cite{Shanahan:2018pib,Hackett:2023rif,Hackett:2023nkr,Wang:2024lrm}
provide a direct, nonperturbative handle on
$\langle T^{\mu}_{\ \mu}\rangle$.
Independent experimental extractions corroborate these values:
(i)~the \emph{quark-sector} GFF is obtained from deep-virtual Compton
scattering at Jefferson Lab \cite{Burkert:2018bqq}, analyzed with the dispersive–factorization approaches of
\cite{Diehl:2007jb,Anikin:2007tx,Pasquini:2014vua};
(ii)~the \emph{gluon-sector} GFF is deduced from near-threshold
$J/\psi$ photoproduction \cite{Duran:2022xag} using both holographic-QCD \cite{Mamo:2019mka,Mamo:2022eui} and an NRQCD–GPD
factorization scheme \cite{Sun:2021gmi}.

\begin{figure}[t]
  \centering
  \includegraphics[width=0.80\linewidth]{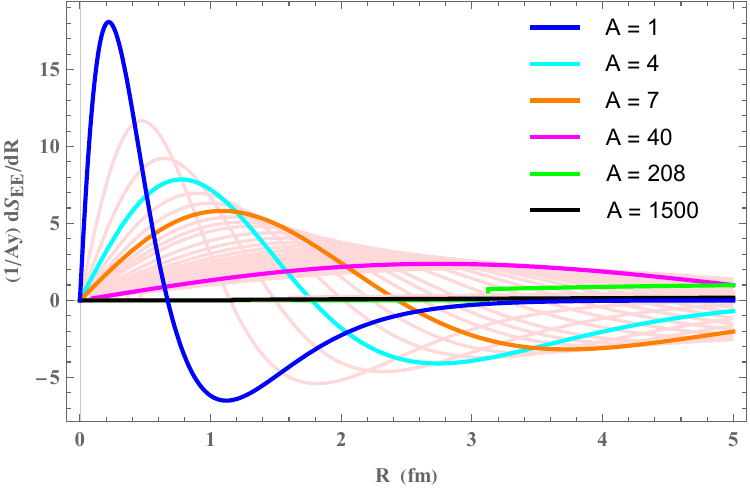}
  \caption{Radial derivative \(\partial_R S_{\!EE}\) for the systems in
  Fig.~\ref{fig:SEE_nuclei}.
  Positive values (\(R<R_{\!EE}\)) represent an outward entropic
  pressure; negative values (\(R>R_{\!EE}\)) pull inward.
  The zero crossing defines the “information wall’’ at \(R_{\!EE}\).}
  \label{fig:DSEE_nuclei}
\end{figure}

The sign-changing entropy gradient (see Fig.~\ref{fig:DSEE_nuclei}) is an intrinsic property of the QCD vacuum, akin to Casimir-like stresses in non-Abelian gauge theories\,\cite{Chernodub:2023dok}, and \emph{not} a by-product of the
$2^{++}$–$0^{++}$ resonances admixture that governs the Breit-frame three-dimensional EMT
pressure inside a hadron\,\cite{Polyakov:2002yz,Mamo:2021krl,cao2025dispersivedeterminationnucleongravitational,Broniowski:2024oyk,broniowski2025gravitationalformfactorsmechanical,Lorce:2025oot} (see Fig.~\ref{fig:pressure} in the Supplemental Material for the separate mechanical-pressure node).

\begin{figure*}[!t]
  \centering
  \subfloat[\(\phi\) photoproduction\label{fig:sigma_phi}]{
    \includegraphics[width=0.35\linewidth]{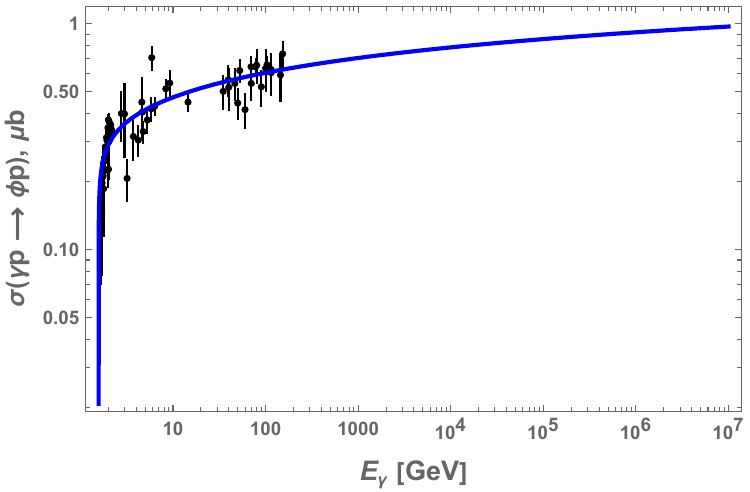}}
  \hspace{0.04\textwidth}
  \subfloat[\(J/\psi\) photoproduction\label{fig:sigma_jpsi}]{\includegraphics[width=0.35\linewidth]{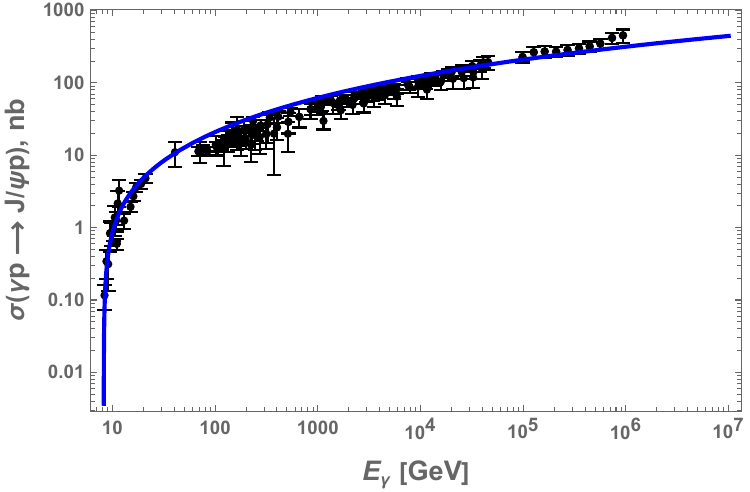}}\\[-0.4ex]
  \subfloat[\(\Upsilon(1S)\) photoproduction\label{fig:sigma_ups}]{
    \includegraphics[width=0.35\linewidth]{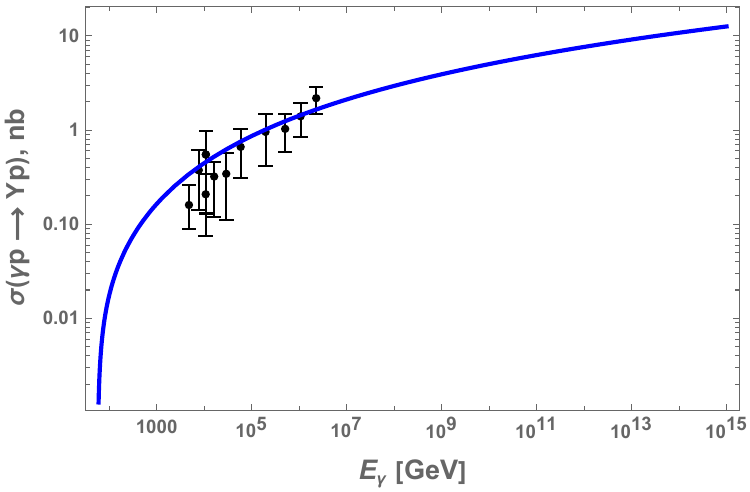}}
  \hspace{0.04\textwidth}
  \subfloat[Elastic \(pp\)\label{fig:sigma_pp}]{
    \includegraphics[width=0.35\linewidth]{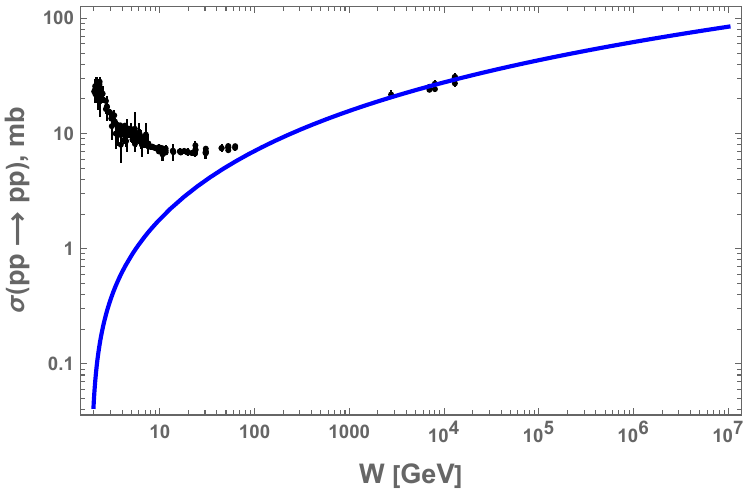}}
 \caption{Cross sections $\sigma$ are plotted versus the rapidity
$y=\arcosh\gamma=\ln\!\bigl(\gamma+\sqrt{\gamma^{2}-1}\bigr)$,
with $\gamma_{\text{beam}}\equiv E_{\gamma}/E_{\gamma,\mathrm{thr}}(V)$
for photoproduction and
$\gamma_{\text{CM}}\equiv \sqrt{s}/(2m_{N})$ for elastic $pp$ scattering. Panels display (a)~\(\phi\) photoproduction, (b)~\(J/\psi\) photoproduction,
(c)~\(\Upsilon(1S)\) photoproduction, and (d)~elastic \(pp\) scattering.
The solid blue curves show the scaling ansatz
\(\sigma \propto y^{\delta}\) with
\(\delta_\phi=0.387\) in~(a) and \(\delta=2\) in~(b)–(d).
The shallower \(y^{0.39}\) rise for \(\phi\) indicates enhanced
infrared sensitivity of light–quark dynamics, whereas heavy quarkonia
and elastic \(pp\) retain the full \(y^{2}\) growth expected from the
high-energy eikonal picture.
Data sources:
\(\phi\) from Refs.~\cite{Egloff:1979mg,Busenitz:1989gq};
\(J/\psi\) and \(\Upsilon\) from the compilation in
Ref.~\cite{Mamo:2019mka};
elastic \(pp\) from
Refs.~\cite{TOTEM:2018hki,ParticleDataGroup:2022pth,ATLAS:2022mgx}.}
  \label{fig:rapidity_scaling}
\end{figure*}

An \emph{entropic ansatz} relates an elastic\,$2\!\to\!2$ cross section
to the mechanical entropy of the color field produced in the
collision\,\cite{Liu:2023zno,Low:2024hvn}; see \textit{Cross sections and data} below for details. Ref.~\cite{Dosch:2023bxj}, following the exponential entropic ansatz of Ref.~\cite{Kharzeev:2017qzs} for DIS structure functions, models total cross sections as $\sigma\propto e^{\delta y}$. Although this form reproduces vector-meson photoproduction data at high energies, it fails near threshold and overshoots the Froissart--Martin unitarity bound in the $y\!\to\!\infty$ limit. By contrast, our power-law ansatz $\sigma\propto y^{\delta}$ describes both threshold and high-energy photoproduction data with a single exponent and remains well below the Froissart--Martin limit across the full kinematic range, as illustrated in Figs.~\ref{fig:sigma_phi}–\ref{fig:sigma_ups}. See also~\cite{Kovner:2015hga,Kovner:2018rbf,Tu:2019ouv,Ramos:2020kaj,Gotsman:2020bjc,H1:2020zpd,Kharzeev:2021yyf,Hentschinski:2021aux,Hentschinski:2022rsa,Hentschinski:2023izh,Hentschinski:2024gaa,Datta:2024hpn,Moriggi:2024tiz} for other proposals which relate entanglement entropy with multiplicity of final-state hadrons produced in high-energy collisions.

In the remainder of this paper we (i) summarize the lattice-QCD inputs that fix \(\rho_h\) and \(R_{\!EE}\); (ii) quantify the entropic force associated with \(\partial_R S_{\!EE}\); and (iii) show how unitarity reduces an entropic ansatz for elastic \(2\!\to\!2\) processes to \(\sigma(s)=\mathcal N^2 y^\delta\), with \(\delta\) reflecting the UV sensitivity of the channel.

\textit{Trace anomaly and lattice inputs.—}
Entropic confinement depends only on the transverse anomaly density
\(\rho_h(b_\perp)=\langle T^{\mu}{}_{\mu}(b_\perp)\rangle\) (two-dimensional impact-parameter density). Modern
lattice data
\cite{Shanahan:2018pib,Pefkou:2021fni,Hackett:2023rif,
       Hackett:2023nkr,Wang:2024lrm}
and perturbative counting rules
\cite{Brodsky:1973kr,Brodsky:1976rz,Brodsky:1980th,Brodsky:1983kb,Tong:2021ctu,Tong:2022zax}
fix the necessary scalar form factors:

\smallskip
\emph{Nucleon ($J=\tfrac12$).—}  With \(t=(P'-P)^2\),
\(\bar P=(P'+P)/2\),
\begin{align}
\langle P'|T_i^{\mu\nu}|P\rangle &=
  \bar u(P')\!\Bigl[
      A_i(t)\gamma^{(\mu}\bar P^{\nu)}
    +D_i(t)\frac{q^\mu q^\nu-g^{\mu\nu}q^{2}}{4m_N}
  \Bigr]u(P), \label{eq:EMT_nucl}
\end{align}
with $i\equiv \text{sea-quark}~(q)~\text{or gluon}~(g)$ and the trace is governed by the dipole
\be
A_i^{S}(t)=A_i(t)-\tfrac{3t}{4m_N^{2}}D_i(t)=
           \alpha_{S,i}(1-t/\Lambda_{S,i}^{2})^{-2}.
\label{eq:AS_nucl}
\ee

\emph{Pion ($J=0$).—} With monopole form
\be
A_\pi^{S}(t)=A_\pi(t)-\tfrac{t}{4m_\pi^{2}}\bigl[2A_\pi(t)+3D_\pi(t)\bigr].
\label{eq:AS_pion}
\ee

\emph{Spin-0 nuclei.—}
With \(n_A=3A-1\),
\be
A_{i}^{S,A}(t)=
  A\,\alpha_{S,i}\,(1-t/\Lambda_{S,i}^{2})^{-n_A}.
\label{eq:AS_nucleus}
\ee

After 2D Fourier transform with respect to $t=-\bm{\Delta}_\perp^2$, any \(n\)-pole form yields the Bessel–$K$ density
\be
\rho_i(b_\perp)=\alpha_{S,i}\frac{\Lambda_{S,i}^{2}}{2\pi}
   \frac{(\Lambda_{S,i}b_\perp)^{n-1}}{2^{n-1}\Gamma(n)}
   K_{n-1}(\Lambda_{S,i}b_\perp).\label{anomlyDensity}
\ee
Summing over \(i=q,g\) and evaluating at \(b_\perp=R_{\!EE}\) fixes
\(c_h=8\pi^{2}R_{\!EE}^{2}\rho_h(R_{\!EE})\) in
Eq.~\eqref{eq:SEE_boost}. All lattice fit parameters are summarized in Table~\ref{tab:gff_params} of the Supplemental Material.

\begin{table}[t]
\caption{Charge radii $r_c$, maximum-entropy radii $R_{\!EE}$, and specific
entropies $S_{\!EE}/(A\,y)$ for representative hadronic and
nuclear systems.  All radii are in~fm.
Proton $r_c$ is from PRad~\cite{Xiong:2019umf};
the pion value is the PDG average~\cite{ParticleDataGroup:2024cfk};
nuclear radii are taken from~\cite{Angeli:2013epw}.
Over the mass range shown the maximum-entropy radius is reproduced to
better than~$0.1\,\%$ by
$ R_{\!EE}(A) = 0.964\,\sqrt{A}\;-\;0.279/\sqrt{A}\;-\;0.014 $.
Note that $R_{\!EE}(A)$ need not match the liquid–drop $r_0A^{1/3}$ law for \emph{geometric} radii because the underlying weights (2D trace anomaly density vs 3D mass density) and the codimension (surface vs volume) are different.}
\begin{ruledtabular}
\begin{tabular}{lccc}
Species & \(r_c\) & \(R_{\!EE}\) & \(S_{\!EE}/(Ay)\)\\\hline
\(\pi^{\pm}\)         & 0.659(4)  & 0.98(6)   & 12(4)  \\
\(N\)                 & 0.831(14) & 0.67(8)   & 7.17(36)\\
\(^4\mathrm{He}\)     & 1.68(3)   & 1.77(21)  & 8.57(43)\\
\(^7\mathrm{Li}\)     & 2.44(4)   & 2.43(28)  & 8.74(44)\\
\(^ {40}\mathrm{Ar}\) & 3.43(3)   & 6.04(71)  & 8.91(45)\\
\(^ {208}\mathrm{Pb}\)& 5.50(1)   & 13.9(1.6) & 8.94(45)
\end{tabular}
\end{ruledtabular}
\label{tab:HadronNucleusEntropy}
\end{table}

\begin{figure}[t]
  \centering
  \includegraphics[width=0.45\textwidth]{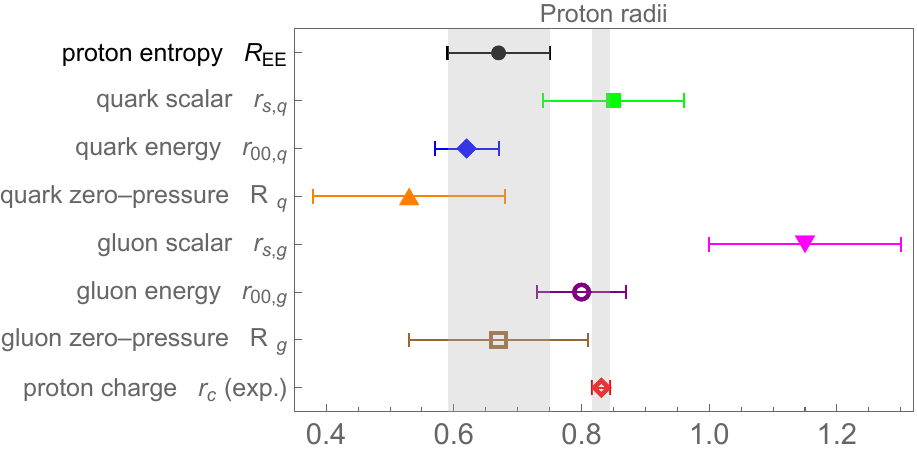}
  \caption{Visual comparison of proton radii using lattice gravitational form factor~\eqref{eq:EMT_nucl} from~\cite{Hackett:2023rif}, see Table~\ref{tab:gff_params}: maximum-entropy radius $R_{\!EE}$, mechanical core $R_{q,g}$~\eqref{eq:Rcore-analytic} (from the node of the 3D Breit-frame pressure density, see Fig.~\ref{fig:pressure}), rms scalar radii $r_s$~\eqref{eq:radii-analytic}, rms energy radii $r_{00}$~\eqref{eq:radii-analytic}, and the experimental rms charge radius $r_c$ from PRad~\cite{Xiong:2019umf}. \emph{All rms radii shown are computed separately for the quark and for the gluon sectors using 3D Breit-frame densities and Eq.~\eqref{eq:radii-analytic} in the Supplemental Material; no $q\!+\!g$ sum is taken. However, $R_{\!EE}$ is computed from $q+g$ 2D (impact-parameter space) trace anomaly density Eq.~\eqref{anomlyDensity}.}\label{fig:radii-comparison}}
\end{figure}

\textit{Mechanical entanglement and confinement.—}
Equation~\eqref{eq:SEE_boost}, together with
Figs.~\ref{fig:SEE_nuclei} and \ref{fig:DSEE_nuclei},
shows that the entropy gradient reverses sign at
$r=R_{\!EE}$ (see~Table~\ref{tab:HadronNucleusEntropy} for values of $R_{\!EE}$ for pion and nuclei, also see Fig.~\ref{fig:radii-comparison} for a visual comparison of various proton radii inferred from the lattice GFFs). Accordingly, the radial derivative \(\partial_{R} S_{\!EE}\) acts as an \emph{entropic pressure or force} in the sense of Refs.\,\cite{Verlinde:2010hp,Fursaev:2010ix}. A parton that strays \emph{inside} the surface
($r<R_{\!EE}$) is pushed outward, whereas one that moves
\emph{outside} ($r>R_{\!EE}$) is pulled back. Colored
degrees of freedom are therefore locked onto the
two-dimensional “information wall’’ at $r=R_{\!EE}$.
This entropic bag complements the Wilson area-law criterion: it provides a high-energy (UV) diagnostic while the area law governs long-distance (IR) dynamics.

\textit{Cross sections and data.—}
The entropic ansatz ties an elastic $2\!\to\!2$ cross section to the entropy of the intermediate color field~\cite{Low:2024hvn}. In the standard eikonal representation, unitarity gives
\be
\sigma_{\mathrm{el}}(s)=\!\int d^{2}b_{\perp}\,\bigl|1-e^{i\chi(b_{\perp},y)}\bigr|^{2},
\ee
and, because the phase accumulates additively with rapidity, we take an entropy-controlled eikonal $\chi(b_{\perp},y)\propto y$. This yields a single-scale power law at high energy. In particular, for small-eikonal (``moderate opacity'') regime, we find $\sigma_{\mathrm{el}}(s)\propto \int d^2b\,|\chi|^2\propto y^2$.

Consequently, for each elastic/exclusive channel, we write the entropic ansatz
\begin{equation}
  \sigma_{\cal C}(s)=\mathcal N_{\cal C}^{2}\,y^{\delta_{\cal C}},\qquad
  y=\arcosh\gamma ,
\label{eq:sigma_entropic}
\end{equation}
where $\mathcal N_{\cal C}$ is a single normalization constant per elastic/exclusive channel $\mathcal{C}$ and the exponent $\delta_{\cal C}$ quantifies how efficiently the process probes the ultraviolet anomaly density on the information wall~\cite{Low:2024hvn}.

\vspace{0.3em}\noindent
\emph{UV–dominated channels.—}
Rapidity is defined as $y=\arcosh\gamma=\ln\bigl(\gamma+\sqrt{\gamma^{2}-1}\bigr)$. For vector-meson photoproduction with $s=W^2=\left(p_{\gamma}+p_p\right)^2$, we adopt the dimensionless beam–energy ratio $\gamma_{\text{beam}}\equiv E_{\gamma}/E_{\gamma,\mathrm{thr}}(V)$, where $E_{\gamma,\mathrm{thr}}(V)=\bigl[(m_{N}+M_{V})^{2}-m_{N}^{2}\bigr]/(2m_{N})$, so that $\gamma_{\text{beam}}=1$ at threshold and $y_{\text{beam}}=\arcosh\gamma_{\text{beam}}$. For elastic $pp$ scattering (with $\sqrt{s}=W$) in the c.m.\ frame the Lorentz factor of either proton is \(\gamma_{\text{CM}}\equiv \sqrt{s}/(2m_{N})\), yielding \(y_{\text{CM}}=\arcosh\gamma_{\text{CM}}\). Reactions characterized by compact transverse sizes
(elastic $pp$, $J/\psi$, and $\Upsilon$ production) sample the full
ultraviolet content of the wall; their exponents saturate at the
unitarity limit $\delta=2$:
\[
\begin{aligned}
\mathcal N_{pp}^{2}       &= 0.324~\text{mb}, &
\mathcal N_{J/\psi}^{2}   &= 2.1\times10^{-6}~\text{mb},\\
\mathcal N_{\Upsilon}^{2} &= 1.3\times10^{-8}~\text{mb}.
\end{aligned}
\]
Even at the largest measured rapidities
($y_{\!\text{CM}}\!\simeq\!10$ at the LHC) the resulting cross sections remain below \(32~\mathrm{mb}\), smaller by a factor \(\gtrsim 2\) than the canonical \(\pi/m_\pi^{2}\simeq 62.8~\mathrm{mb}\) scale;
see Figs.~\ref{fig:sigma_jpsi}–\ref{fig:sigma_pp}.

\vspace{0.3em}\noindent
\emph{IR-sensitive channel.—}
The more diffuse $\phi$ meson samples the wall less efficiently,
yielding a softer rise
\[
  \delta_{\phi}=0.387,\qquad
  \mathcal N_{\phi}^{2}=3.3\times10^{-4}~\text{mb},
\]
so that $\sigma_{\gamma p\to\phi p}$ stays far below
\(\sigma_{\!_{\mathrm{FM}}}\) across the entire energy range
from threshold to multi-TeV photoproduction; see Fig.~\ref{fig:sigma_phi}.

\vspace{0.3em}\noindent
All extracted cross sections remain well below the canonical Froissart--Martin scale \((\pi/m_\pi^{2})\ln^{2}\!s\) for \(\sigma_{\mathrm{tot}}\) (with \(\pi/m_\pi^{2}\simeq 62.8~\mathrm{mb}\) after unit conversion): elastic \(pp\) lies by factors \(\gtrsim 2\) below at present energies, whereas heavy-quarkonia photoproduction is suppressed by \(\sim 10\!-\!10^{3}\).

\textit{Conclusions.—}
We have shown that the QCD trace anomaly, quantified with
percent-level lattice determinations of the scalar gravitational form
factors, fixes a transverse ``information wall’’ of radius
\(R_{\!EE}\). Tracing over quark and gluon modes outside that surface
yields a finite, \emph{parameter-free} mechanical entropy
\(S_{\!EE}=c_h\,y\) with
\emph{$c_h$ fixed entirely by lattice inputs}—whose sign-changing radial derivative locks colored
fields to the wall at \emph{all} momentum scales. This entropic mechanism provides a high-energy diagnostic that \emph{complements} the traditional Wilson-loop area law: it remains consistent with asymptotic freedom in the ultraviolet and with linear confinement in the infrared. Imposing unitarity on an entropic ansatz for the amplitude yields
\(\sigma(s)\propto y^{\delta}\). Hard processes
(\(pp\), \(J/\psi\), \(\Upsilon\)) saturate the unitarity limit \(\delta=2\), exhibiting the familiar \(y^{2}\!\sim\!\ln^{2}\!s\) growth pattern, while the more diffuse \(\phi\) channel selects \(\delta_{\phi}=0.387\). All extracted cross sections remain well below the canonical Froissart--Martin bound \((\pi/m_{\pi}^{2})\ln^{2}\!s\) for \(\sigma_{\mathrm{tot}}\).

\textit{Outlook.—}
The entropic framework developed here delivers a quantitative baseline against which future high-precision data can be benchmarked. Upcoming measurements at the HL-LHC, Jefferson Lab (CLAS12, GlueX, and SoLID), and the Electron-Ion Collider will confront its predictions across an unprecedented kinematic range. Any systematic deviation---for example from odderon exchange, Regge exchanges, exotic multiquark resonances---will stand out cleanly against the fully quantified entropy background, offering a sharpened path to discovering genuinely new QCD dynamics.

\begin{acknowledgments}
K.M.\ was supported by the U.S.\ Department of Energy under Grant No.\ DE-FG02-04ER41302.
Additional support for this work was provided by the DOE Office of Science through Contract No.\ DE-AC05-06OR23177, under which Jefferson Science Associates, LLC operates the Thomas Jefferson National Accelerator Facility.
\end{acknowledgments}

\section*{DATA AVAILABILITY}
The data that support the findings of this article are not publicly available. The data are available from the authors upon reasonable request.

\bibliography{entropic_confinement_refs}

\begin{thebibliography}{92}%
\makeatletter
\providecommand \@ifxundefined [1]{%
 \@ifx{#1\undefined}
}%
\providecommand \@ifnum [1]{%
 \ifnum #1\expandafter \@firstoftwo
 \else \expandafter \@secondoftwo
 \fi
}%
\providecommand \@ifx [1]{%
 \ifx #1\expandafter \@firstoftwo
 \else \expandafter \@secondoftwo
 \fi
}%
\providecommand \natexlab [1]{#1}%
\providecommand \enquote  [1]{``#1''}%
\providecommand \bibnamefont  [1]{#1}%
\providecommand \bibfnamefont [1]{#1}%
\providecommand \citenamefont [1]{#1}%
\providecommand \href@noop [0]{\@secondoftwo}%
\providecommand \href [0]{\begingroup \@sanitize@url \@href}%
\providecommand \@href[1]{\@@startlink{#1}\@@href}%
\providecommand \@@href[1]{\endgroup#1\@@endlink}%
\providecommand \@sanitize@url [0]{\catcode `\\12\catcode `\$12\catcode `\&12\catcode `\#12\catcode `\^12\catcode `\_12\catcode `\%12\relax}%
\providecommand \@@startlink[1]{}%
\providecommand \@@endlink[0]{}%
\providecommand \url  [0]{\begingroup\@sanitize@url \@url }%
\providecommand \@url [1]{\endgroup\@href {#1}{\urlprefix }}%
\providecommand \urlprefix  [0]{URL }%
\providecommand \Eprint [0]{\href }%
\providecommand \doibase [0]{https://doi.org/}%
\providecommand \selectlanguage [0]{\@gobble}%
\providecommand \bibinfo  [0]{\@secondoftwo}%
\providecommand \bibfield  [0]{\@secondoftwo}%
\providecommand \translation [1]{[#1]}%
\providecommand \BibitemOpen [0]{}%
\providecommand \bibitemStop [0]{}%
\providecommand \bibitemNoStop [0]{.\EOS\space}%
\providecommand \EOS [0]{\spacefactor3000\relax}%
\providecommand \BibitemShut  [1]{\csname bibitem#1\endcsname}%
\let\auto@bib@innerbib\@empty
\bibitem [{\citenamefont {Wilson}(1974)}]{Wilson:1974sk}%
  \BibitemOpen
  \bibfield  {author} {\bibinfo {author} {\bibfnamefont {K.~G.}\ \bibnamefont {Wilson}},\ }\bibfield  {title} {\bibinfo {title} {{Confinement of Quarks}},\ }\href {https://doi.org/10.1103/PhysRevD.10.2445} {\bibfield  {journal} {\bibinfo  {journal} {Phys. Rev. D}\ }\textbf {\bibinfo {volume} {10}},\ \bibinfo {pages} {2445} (\bibinfo {year} {1974})}\BibitemShut {NoStop}%
\bibitem [{\citenamefont {Rosenhaus}\ and\ \citenamefont {Smolkin}(2014{\natexlab{a}})}]{Rosenhaus:2014nha}%
  \BibitemOpen
  \bibfield  {author} {\bibinfo {author} {\bibfnamefont {V.}~\bibnamefont {Rosenhaus}}\ and\ \bibinfo {author} {\bibfnamefont {M.}~\bibnamefont {Smolkin}},\ }\bibfield  {title} {\bibinfo {title} {{Entanglement Entropy Flow and the Ward Identity}},\ }\href {https://doi.org/10.1103/PhysRevLett.113.261602} {\bibfield  {journal} {\bibinfo  {journal} {Phys. Rev. Lett.}\ }\textbf {\bibinfo {volume} {113}},\ \bibinfo {pages} {261602} (\bibinfo {year} {2014}{\natexlab{a}})},\ \Eprint {https://arxiv.org/abs/1406.2716} {arXiv:1406.2716 [hep-th]} \BibitemShut {NoStop}%
\bibitem [{\citenamefont {Ben-Ami}\ \emph {et~al.}(2015)\citenamefont {Ben-Ami}, \citenamefont {Carmi},\ and\ \citenamefont {Smolkin}}]{Ben-Ami:2015zsa}%
  \BibitemOpen
  \bibfield  {author} {\bibinfo {author} {\bibfnamefont {O.}~\bibnamefont {Ben-Ami}}, \bibinfo {author} {\bibfnamefont {D.}~\bibnamefont {Carmi}},\ and\ \bibinfo {author} {\bibfnamefont {M.}~\bibnamefont {Smolkin}},\ }\bibfield  {title} {\bibinfo {title} {{Renormalization group flow of entanglement entropy on spheres}},\ }\href {https://doi.org/10.1007/JHEP08(2015)048} {\bibfield  {journal} {\bibinfo  {journal} {JHEP}\ }\textbf {\bibinfo {volume} {08}},\ \bibinfo {pages} {048}},\ \Eprint {https://arxiv.org/abs/1504.00913} {arXiv:1504.00913 [hep-th]} \BibitemShut {NoStop}%
\bibitem [{\citenamefont {Bombelli}\ \emph {et~al.}(1986)\citenamefont {Bombelli}, \citenamefont {Koul}, \citenamefont {Lee},\ and\ \citenamefont {Sorkin}}]{Bombelli:1986rw}%
  \BibitemOpen
  \bibfield  {author} {\bibinfo {author} {\bibfnamefont {L.}~\bibnamefont {Bombelli}}, \bibinfo {author} {\bibfnamefont {R.~K.}\ \bibnamefont {Koul}}, \bibinfo {author} {\bibfnamefont {J.}~\bibnamefont {Lee}},\ and\ \bibinfo {author} {\bibfnamefont {R.~D.}\ \bibnamefont {Sorkin}},\ }\bibfield  {title} {\bibinfo {title} {{A quantum source of entropy for black holes}},\ }\href {https://doi.org/10.1103/PhysRevD.34.373} {\bibfield  {journal} {\bibinfo  {journal} {Phys. Rev. D}\ }\textbf {\bibinfo {volume} {34}},\ \bibinfo {pages} {373} (\bibinfo {year} {1986})}\BibitemShut {NoStop}%
\bibitem [{\citenamefont {Srednicki}(1993)}]{Srednicki:1993im}%
  \BibitemOpen
  \bibfield  {author} {\bibinfo {author} {\bibfnamefont {M.}~\bibnamefont {Srednicki}},\ }\bibfield  {title} {\bibinfo {title} {{Entropy and area}},\ }\href {https://doi.org/10.1103/PhysRevLett.71.666} {\bibfield  {journal} {\bibinfo  {journal} {Phys. Rev. Lett.}\ }\textbf {\bibinfo {volume} {71}},\ \bibinfo {pages} {666} (\bibinfo {year} {1993})},\ \Eprint {https://arxiv.org/abs/hep-th/9303048} {arXiv:hep-th/9303048} \BibitemShut {NoStop}%
\bibitem [{\citenamefont {Kabat}\ and\ \citenamefont {Strassler}(1994)}]{Kabat:1994vj}%
  \BibitemOpen
  \bibfield  {author} {\bibinfo {author} {\bibfnamefont {D.~N.}\ \bibnamefont {Kabat}}\ and\ \bibinfo {author} {\bibfnamefont {M.~J.}\ \bibnamefont {Strassler}},\ }\bibfield  {title} {\bibinfo {title} {{A Comment on entropy and area}},\ }\href {https://doi.org/10.1016/0370-2693(94)90515-0} {\bibfield  {journal} {\bibinfo  {journal} {Phys. Lett. B}\ }\textbf {\bibinfo {volume} {329}},\ \bibinfo {pages} {46} (\bibinfo {year} {1994})},\ \Eprint {https://arxiv.org/abs/hep-th/9401125} {arXiv:hep-th/9401125} \BibitemShut {NoStop}%
\bibitem [{\citenamefont {Casini}\ and\ \citenamefont {Huerta}(2007)}]{Casini:2006es}%
  \BibitemOpen
  \bibfield  {author} {\bibinfo {author} {\bibfnamefont {H.}~\bibnamefont {Casini}}\ and\ \bibinfo {author} {\bibfnamefont {M.}~\bibnamefont {Huerta}},\ }\bibfield  {title} {\bibinfo {title} {{A c-theorem for the entanglement entropy}},\ }\href {https://doi.org/10.1088/1751-8113/40/25/S57} {\bibfield  {journal} {\bibinfo  {journal} {J. Phys. A}\ }\textbf {\bibinfo {volume} {40}},\ \bibinfo {pages} {7031} (\bibinfo {year} {2007})},\ \Eprint {https://arxiv.org/abs/cond-mat/0610375} {arXiv:cond-mat/0610375} \BibitemShut {NoStop}%
\bibitem [{\citenamefont {Ryu}\ and\ \citenamefont {Takayanagi}(2006{\natexlab{a}})}]{Ryu:2006bv}%
  \BibitemOpen
  \bibfield  {author} {\bibinfo {author} {\bibfnamefont {S.}~\bibnamefont {Ryu}}\ and\ \bibinfo {author} {\bibfnamefont {T.}~\bibnamefont {Takayanagi}},\ }\bibfield  {title} {\bibinfo {title} {{Holographic derivation of entanglement entropy from AdS/CFT}},\ }\href {https://doi.org/10.1103/PhysRevLett.96.181602} {\bibfield  {journal} {\bibinfo  {journal} {Phys. Rev. Lett.}\ }\textbf {\bibinfo {volume} {96}},\ \bibinfo {pages} {181602} (\bibinfo {year} {2006}{\natexlab{a}})},\ \Eprint {https://arxiv.org/abs/hep-th/0603001} {arXiv:hep-th/0603001} \BibitemShut {NoStop}%
\bibitem [{\citenamefont {Ryu}\ and\ \citenamefont {Takayanagi}(2006{\natexlab{b}})}]{Ryu:2006ef}%
  \BibitemOpen
  \bibfield  {author} {\bibinfo {author} {\bibfnamefont {S.}~\bibnamefont {Ryu}}\ and\ \bibinfo {author} {\bibfnamefont {T.}~\bibnamefont {Takayanagi}},\ }\bibfield  {title} {\bibinfo {title} {{Aspects of Holographic Entanglement Entropy}},\ }\href {https://doi.org/10.1088/1126-6708/2006/08/045} {\bibfield  {journal} {\bibinfo  {journal} {JHEP}\ }\textbf {\bibinfo {volume} {08}},\ \bibinfo {pages} {045}},\ \Eprint {https://arxiv.org/abs/hep-th/0605073} {arXiv:hep-th/0605073} \BibitemShut {NoStop}%
\bibitem [{\citenamefont {Klebanov}\ \emph {et~al.}(2008)\citenamefont {Klebanov}, \citenamefont {Kutasov},\ and\ \citenamefont {Murugan}}]{Klebanov:2007ws}%
  \BibitemOpen
  \bibfield  {author} {\bibinfo {author} {\bibfnamefont {I.~R.}\ \bibnamefont {Klebanov}}, \bibinfo {author} {\bibfnamefont {D.}~\bibnamefont {Kutasov}},\ and\ \bibinfo {author} {\bibfnamefont {A.}~\bibnamefont {Murugan}},\ }\bibfield  {title} {\bibinfo {title} {{Entanglement as a probe of confinement}},\ }\href {https://doi.org/10.1016/j.nuclphysb.2007.12.017} {\bibfield  {journal} {\bibinfo  {journal} {Nucl. Phys. B}\ }\textbf {\bibinfo {volume} {796}},\ \bibinfo {pages} {274} (\bibinfo {year} {2008})},\ \Eprint {https://arxiv.org/abs/0709.2140} {arXiv:0709.2140 [hep-th]} \BibitemShut {NoStop}%
\bibitem [{\citenamefont {Solodukhin}(2008)}]{Solodukhin:2008dh}%
  \BibitemOpen
  \bibfield  {author} {\bibinfo {author} {\bibfnamefont {S.~N.}\ \bibnamefont {Solodukhin}},\ }\bibfield  {title} {\bibinfo {title} {{Entanglement entropy, conformal invariance and extrinsic geometry}},\ }\href {https://doi.org/10.1016/j.physletb.2008.05.071} {\bibfield  {journal} {\bibinfo  {journal} {Phys. Lett. B}\ }\textbf {\bibinfo {volume} {665}},\ \bibinfo {pages} {305} (\bibinfo {year} {2008})},\ \Eprint {https://arxiv.org/abs/0802.3117} {arXiv:0802.3117} \BibitemShut {NoStop}%
\bibitem [{\citenamefont {Casini}\ and\ \citenamefont {Huerta}(2009)}]{Casini:2009sr}%
  \BibitemOpen
  \bibfield  {author} {\bibinfo {author} {\bibfnamefont {H.}~\bibnamefont {Casini}}\ and\ \bibinfo {author} {\bibfnamefont {M.}~\bibnamefont {Huerta}},\ }\bibfield  {title} {\bibinfo {title} {{Entanglement entropy in free quantum field theory}},\ }\href {https://doi.org/10.1088/1126-6708/2009/09/013} {\bibfield  {journal} {\bibinfo  {journal} {JHEP}\ }\textbf {\bibinfo {volume} {09}},\ \bibinfo {pages} {013}},\ \Eprint {https://arxiv.org/abs/0905.2562} {arXiv:0905.2562} \BibitemShut {NoStop}%
\bibitem [{\citenamefont {Casini}\ \emph {et~al.}(2011)\citenamefont {Casini}, \citenamefont {Huerta},\ and\ \citenamefont {Myers}}]{Casini:2011kv}%
  \BibitemOpen
  \bibfield  {author} {\bibinfo {author} {\bibfnamefont {H.}~\bibnamefont {Casini}}, \bibinfo {author} {\bibfnamefont {M.}~\bibnamefont {Huerta}},\ and\ \bibinfo {author} {\bibfnamefont {R.~C.}\ \bibnamefont {Myers}},\ }\bibfield  {title} {\bibinfo {title} {{Towards a derivation of holographic entanglement entropy}},\ }\href {https://doi.org/10.1007/JHEP05(2011)036} {\bibfield  {journal} {\bibinfo  {journal} {JHEP}\ }\textbf {\bibinfo {volume} {05}},\ \bibinfo {pages} {036}},\ \Eprint {https://arxiv.org/abs/1102.0440} {arXiv:1102.0440 [hep-th]} \BibitemShut {NoStop}%
\bibitem [{\citenamefont {Rosenhaus}\ and\ \citenamefont {Smolkin}(2014{\natexlab{b}})}]{Rosenhaus:2014woa}%
  \BibitemOpen
  \bibfield  {author} {\bibinfo {author} {\bibfnamefont {V.}~\bibnamefont {Rosenhaus}}\ and\ \bibinfo {author} {\bibfnamefont {M.}~\bibnamefont {Smolkin}},\ }\bibfield  {title} {\bibinfo {title} {{Entanglement Entropy: A Perturbative Calculation}},\ }\href {https://doi.org/10.1007/JHEP12(2014)179} {\bibfield  {journal} {\bibinfo  {journal} {JHEP}\ }\textbf {\bibinfo {volume} {12}},\ \bibinfo {pages} {179}},\ \Eprint {https://arxiv.org/abs/1403.3733} {arXiv:1403.3733 [hep-th]} \BibitemShut {NoStop}%
\bibitem [{\citenamefont {Soper}(1977)}]{Soper:1976jc}%
  \BibitemOpen
  \bibfield  {author} {\bibinfo {author} {\bibfnamefont {D.~E.}\ \bibnamefont {Soper}},\ }\bibfield  {title} {\bibinfo {title} {{The Parton Model and the Bethe-Salpeter Wave Function}},\ }\href {https://doi.org/10.1103/PhysRevD.15.1141} {\bibfield  {journal} {\bibinfo  {journal} {Phys. Rev. D}\ }\textbf {\bibinfo {volume} {15}},\ \bibinfo {pages} {1141} (\bibinfo {year} {1977})}\BibitemShut {NoStop}%
\bibitem [{\citenamefont {Burkardt}(2000)}]{Burkardt:2000za}%
  \BibitemOpen
  \bibfield  {author} {\bibinfo {author} {\bibfnamefont {M.}~\bibnamefont {Burkardt}},\ }\bibfield  {title} {\bibinfo {title} {{Impact parameter dependent parton distributions and off forward parton distributions for zeta ---\ensuremath{>} 0}},\ }\href {https://doi.org/10.1103/PhysRevD.62.071503} {\bibfield  {journal} {\bibinfo  {journal} {Phys. Rev. D}\ }\textbf {\bibinfo {volume} {62}},\ \bibinfo {pages} {071503} (\bibinfo {year} {2000})},\ \bibinfo {note} {[Erratum: Phys.Rev.D 66, 119903 (2002)]},\ \Eprint {https://arxiv.org/abs/hep-ph/0005108} {arXiv:hep-ph/0005108} \BibitemShut {NoStop}%
\bibitem [{\citenamefont {Stoffers}\ and\ \citenamefont {Zahed}(2013)}]{Stoffers:2012mn}%
  \BibitemOpen
  \bibfield  {author} {\bibinfo {author} {\bibfnamefont {A.}~\bibnamefont {Stoffers}}\ and\ \bibinfo {author} {\bibfnamefont {I.}~\bibnamefont {Zahed}},\ }\bibfield  {title} {\bibinfo {title} {{Holographic Pomeron and entropy}},\ }\href {https://doi.org/10.1103/PhysRevD.88.025038} {\bibfield  {journal} {\bibinfo  {journal} {Phys. Rev. D}\ }\textbf {\bibinfo {volume} {88}},\ \bibinfo {pages} {025038} (\bibinfo {year} {2013})},\ \Eprint {https://arxiv.org/abs/1211.3077} {arXiv:1211.3077} \BibitemShut {NoStop}%
\bibitem [{\citenamefont {Liu}\ and\ \citenamefont {Zahed}(2019)}]{Liu:2018gae}%
  \BibitemOpen
  \bibfield  {author} {\bibinfo {author} {\bibfnamefont {Y.}~\bibnamefont {Liu}}\ and\ \bibinfo {author} {\bibfnamefont {I.}~\bibnamefont {Zahed}},\ }\bibfield  {title} {\bibinfo {title} {{Entanglement in Regge scattering using the AdS/CFT correspondence}},\ }\href {https://doi.org/10.1103/PhysRevD.100.046005} {\bibfield  {journal} {\bibinfo  {journal} {Phys. Rev. D}\ }\textbf {\bibinfo {volume} {100}},\ \bibinfo {pages} {046005} (\bibinfo {year} {2019})},\ \Eprint {https://arxiv.org/abs/1803.09157} {arXiv:1803.09157} \BibitemShut {NoStop}%
\bibitem [{\citenamefont {Kharzeev}\ and\ \citenamefont {Levin}(2017)}]{Kharzeev:2017qzs}%
  \BibitemOpen
  \bibfield  {author} {\bibinfo {author} {\bibfnamefont {D.~E.}\ \bibnamefont {Kharzeev}}\ and\ \bibinfo {author} {\bibfnamefont {E.~M.}\ \bibnamefont {Levin}},\ }\bibfield  {title} {\bibinfo {title} {{Deep inelastic scattering as a probe of entanglement}},\ }\href {https://doi.org/10.1103/PhysRevD.95.114008} {\bibfield  {journal} {\bibinfo  {journal} {Phys. Rev. D}\ }\textbf {\bibinfo {volume} {95}},\ \bibinfo {pages} {114008} (\bibinfo {year} {2017})},\ \Eprint {https://arxiv.org/abs/1702.03489} {arXiv:1702.03489} \BibitemShut {NoStop}%
\bibitem [{\citenamefont {G{\"u}rsoy}\ \emph {et~al.}(2024)\citenamefont {G{\"u}rsoy}, \citenamefont {Kharzeev},\ and\ \citenamefont {Pedraza}}]{Gursoy:2023hge}%
  \BibitemOpen
  \bibfield  {author} {\bibinfo {author} {\bibfnamefont {U.}~\bibnamefont {G{\"u}rsoy}}, \bibinfo {author} {\bibfnamefont {D.~E.}\ \bibnamefont {Kharzeev}},\ and\ \bibinfo {author} {\bibfnamefont {J.~F.}\ \bibnamefont {Pedraza}},\ }\bibfield  {title} {\bibinfo {title} {{Universal rapidity scaling of entanglement entropy inside hadrons from conformal invariance}},\ }\href {https://doi.org/10.1103/PhysRevD.110.074008} {\bibfield  {journal} {\bibinfo  {journal} {Phys. Rev. D}\ }\textbf {\bibinfo {volume} {110}},\ \bibinfo {pages} {074008} (\bibinfo {year} {2024})},\ \Eprint {https://arxiv.org/abs/2306.16145} {arXiv:2306.16145 [hep-th]} \BibitemShut {NoStop}%
\bibitem [{\citenamefont {Kuraev}\ \emph {et~al.}(1977)\citenamefont {Kuraev}, \citenamefont {Lipatov},\ and\ \citenamefont {Fadin}}]{Kuraev:1977fs}%
  \BibitemOpen
  \bibfield  {author} {\bibinfo {author} {\bibfnamefont {E.~A.}\ \bibnamefont {Kuraev}}, \bibinfo {author} {\bibfnamefont {L.~N.}\ \bibnamefont {Lipatov}},\ and\ \bibinfo {author} {\bibfnamefont {V.~S.}\ \bibnamefont {Fadin}},\ }\bibfield  {title} {\bibinfo {title} {{The pomeranchuk singularity in nonabelian gauge theories}},\ }\href@noop {} {\bibfield  {journal} {\bibinfo  {journal} {Sov. Phys. JETP}\ }\textbf {\bibinfo {volume} {45}},\ \bibinfo {pages} {199} (\bibinfo {year} {1977})}\BibitemShut {NoStop}%
\bibitem [{\citenamefont {Balitsky}\ and\ \citenamefont {Lipatov}(1978)}]{Balitsky:1978ic}%
  \BibitemOpen
  \bibfield  {author} {\bibinfo {author} {\bibfnamefont {I.}~\bibnamefont {Balitsky}}\ and\ \bibinfo {author} {\bibfnamefont {L.~N.}\ \bibnamefont {Lipatov}},\ }\bibfield  {title} {\bibinfo {title} {{The Pomeranchuk singularity in quantum chromodynamics}},\ }\href@noop {} {\bibfield  {journal} {\bibinfo  {journal} {Sov. J. Nucl. Phys.}\ }\textbf {\bibinfo {volume} {28}},\ \bibinfo {pages} {822} (\bibinfo {year} {1978})}\BibitemShut {NoStop}%
\bibitem [{\citenamefont {Lipatov}(1997)}]{Lipatov:1996ts}%
  \BibitemOpen
  \bibfield  {author} {\bibinfo {author} {\bibfnamefont {L.~N.}\ \bibnamefont {Lipatov}},\ }\bibfield  {title} {\bibinfo {title} {{Small-x physics in perturbative QCD}},\ }\href {https://doi.org/10.1016/S0370-1573(96)00045-2} {\bibfield  {journal} {\bibinfo  {journal} {Phys. Rept.}\ }\textbf {\bibinfo {volume} {286}},\ \bibinfo {pages} {131} (\bibinfo {year} {1997})},\ \Eprint {https://arxiv.org/abs/hep-ph/9610276} {arXiv:hep-ph/9610276} \BibitemShut {NoStop}%
\bibitem [{\citenamefont {Jalilian-Marian}\ \emph {et~al.}(1997)\citenamefont {Jalilian-Marian}, \citenamefont {Kovner}, \citenamefont {Leonidov},\ and\ \citenamefont {Weigert}}]{Jalilian-Marian:1997qno}%
  \BibitemOpen
  \bibfield  {author} {\bibinfo {author} {\bibfnamefont {J.}~\bibnamefont {Jalilian-Marian}}, \bibinfo {author} {\bibfnamefont {A.}~\bibnamefont {Kovner}}, \bibinfo {author} {\bibfnamefont {A.}~\bibnamefont {Leonidov}},\ and\ \bibinfo {author} {\bibfnamefont {H.}~\bibnamefont {Weigert}},\ }\bibfield  {title} {\bibinfo {title} {{The BFKL equation from the Wilson renormalization group}},\ }\href {https://doi.org/10.1016/S0550-3213(97)00440-9} {\bibfield  {journal} {\bibinfo  {journal} {Nucl. Phys. B}\ }\textbf {\bibinfo {volume} {504}},\ \bibinfo {pages} {415} (\bibinfo {year} {1997})},\ \Eprint {https://arxiv.org/abs/hep-ph/9701284} {arXiv:hep-ph/9701284} \BibitemShut {NoStop}%
\bibitem [{\citenamefont {Jalilian-Marian}\ \emph {et~al.}(1998{\natexlab{a}})\citenamefont {Jalilian-Marian}, \citenamefont {Kovner}, \citenamefont {Leonidov},\ and\ \citenamefont {Weigert}}]{Jalilian-Marian:1997jhx}%
  \BibitemOpen
  \bibfield  {author} {\bibinfo {author} {\bibfnamefont {J.}~\bibnamefont {Jalilian-Marian}}, \bibinfo {author} {\bibfnamefont {A.}~\bibnamefont {Kovner}}, \bibinfo {author} {\bibfnamefont {A.}~\bibnamefont {Leonidov}},\ and\ \bibinfo {author} {\bibfnamefont {H.}~\bibnamefont {Weigert}},\ }\bibfield  {title} {\bibinfo {title} {{The Wilson renormalization group for low x physics: Towards the high density regime}},\ }\href {https://doi.org/10.1103/PhysRevD.59.014014} {\bibfield  {journal} {\bibinfo  {journal} {Phys. Rev. D}\ }\textbf {\bibinfo {volume} {59}},\ \bibinfo {pages} {014014} (\bibinfo {year} {1998}{\natexlab{a}})},\ \Eprint {https://arxiv.org/abs/hep-ph/9706377} {arXiv:hep-ph/9706377} \BibitemShut {NoStop}%
\bibitem [{\citenamefont {Jalilian-Marian}\ \emph {et~al.}(1998{\natexlab{b}})\citenamefont {Jalilian-Marian}, \citenamefont {Kovner},\ and\ \citenamefont {Weigert}}]{Jalilian-Marian:1997ubg}%
  \BibitemOpen
  \bibfield  {author} {\bibinfo {author} {\bibfnamefont {J.}~\bibnamefont {Jalilian-Marian}}, \bibinfo {author} {\bibfnamefont {A.}~\bibnamefont {Kovner}},\ and\ \bibinfo {author} {\bibfnamefont {H.}~\bibnamefont {Weigert}},\ }\bibfield  {title} {\bibinfo {title} {{The Wilson renormalization group for low x physics: Gluon evolution at finite parton density}},\ }\href {https://doi.org/10.1103/PhysRevD.59.014015} {\bibfield  {journal} {\bibinfo  {journal} {Phys. Rev. D}\ }\textbf {\bibinfo {volume} {59}},\ \bibinfo {pages} {014015} (\bibinfo {year} {1998}{\natexlab{b}})},\ \Eprint {https://arxiv.org/abs/hep-ph/9709432} {arXiv:hep-ph/9709432} \BibitemShut {NoStop}%
\bibitem [{\citenamefont {Kovner}\ and\ \citenamefont {Milhano}(2000)}]{Kovner:1999bj}%
  \BibitemOpen
  \bibfield  {author} {\bibinfo {author} {\bibfnamefont {A.}~\bibnamefont {Kovner}}\ and\ \bibinfo {author} {\bibfnamefont {J.~G.}\ \bibnamefont {Milhano}},\ }\bibfield  {title} {\bibinfo {title} {{Vector potential versus color charge density in low x evolution}},\ }\href {https://doi.org/10.1103/PhysRevD.61.014012} {\bibfield  {journal} {\bibinfo  {journal} {Phys. Rev. D}\ }\textbf {\bibinfo {volume} {61}},\ \bibinfo {pages} {014012} (\bibinfo {year} {2000})},\ \Eprint {https://arxiv.org/abs/hep-ph/9904420} {arXiv:hep-ph/9904420} \BibitemShut {NoStop}%
\bibitem [{\citenamefont {Kovner}\ \emph {et~al.}(2000)\citenamefont {Kovner}, \citenamefont {Milhano},\ and\ \citenamefont {Weigert}}]{Kovner:2000pt}%
  \BibitemOpen
  \bibfield  {author} {\bibinfo {author} {\bibfnamefont {A.}~\bibnamefont {Kovner}}, \bibinfo {author} {\bibfnamefont {J.~G.}\ \bibnamefont {Milhano}},\ and\ \bibinfo {author} {\bibfnamefont {H.}~\bibnamefont {Weigert}},\ }\bibfield  {title} {\bibinfo {title} {{Relating different approaches to nonlinear QCD evolution at finite gluon density}},\ }\href {https://doi.org/10.1103/PhysRevD.62.114005} {\bibfield  {journal} {\bibinfo  {journal} {Phys. Rev. D}\ }\textbf {\bibinfo {volume} {62}},\ \bibinfo {pages} {114005} (\bibinfo {year} {2000})},\ \Eprint {https://arxiv.org/abs/hep-ph/0004014} {arXiv:hep-ph/0004014} \BibitemShut {NoStop}%
\bibitem [{\citenamefont {Weigert}(2002)}]{Weigert:2000gi}%
  \BibitemOpen
  \bibfield  {author} {\bibinfo {author} {\bibfnamefont {H.}~\bibnamefont {Weigert}},\ }\bibfield  {title} {\bibinfo {title} {{Unitarity at small Bjorken x}},\ }\href {https://doi.org/10.1016/S0375-9474(01)01668-2} {\bibfield  {journal} {\bibinfo  {journal} {Nucl. Phys. A}\ }\textbf {\bibinfo {volume} {703}},\ \bibinfo {pages} {823} (\bibinfo {year} {2002})},\ \Eprint {https://arxiv.org/abs/hep-ph/0004044} {arXiv:hep-ph/0004044} \BibitemShut {NoStop}%
\bibitem [{\citenamefont {Iancu}\ \emph {et~al.}(2001{\natexlab{a}})\citenamefont {Iancu}, \citenamefont {Leonidov},\ and\ \citenamefont {McLerran}}]{Iancu:2000hn}%
  \BibitemOpen
  \bibfield  {author} {\bibinfo {author} {\bibfnamefont {E.}~\bibnamefont {Iancu}}, \bibinfo {author} {\bibfnamefont {A.}~\bibnamefont {Leonidov}},\ and\ \bibinfo {author} {\bibfnamefont {L.~D.}\ \bibnamefont {McLerran}},\ }\bibfield  {title} {\bibinfo {title} {{Nonlinear gluon evolution in the color glass condensate. 1.}},\ }\href {https://doi.org/10.1016/S0375-9474(01)00642-X} {\bibfield  {journal} {\bibinfo  {journal} {Nucl. Phys. A}\ }\textbf {\bibinfo {volume} {692}},\ \bibinfo {pages} {583} (\bibinfo {year} {2001}{\natexlab{a}})},\ \Eprint {https://arxiv.org/abs/hep-ph/0011241} {arXiv:hep-ph/0011241} \BibitemShut {NoStop}%
\bibitem [{\citenamefont {Iancu}\ \emph {et~al.}(2001{\natexlab{b}})\citenamefont {Iancu}, \citenamefont {Leonidov},\ and\ \citenamefont {McLerran}}]{Iancu:2001ad}%
  \BibitemOpen
  \bibfield  {author} {\bibinfo {author} {\bibfnamefont {E.}~\bibnamefont {Iancu}}, \bibinfo {author} {\bibfnamefont {A.}~\bibnamefont {Leonidov}},\ and\ \bibinfo {author} {\bibfnamefont {L.~D.}\ \bibnamefont {McLerran}},\ }\bibfield  {title} {\bibinfo {title} {{The Renormalization group equation for the color glass condensate}},\ }\href {https://doi.org/10.1016/S0370-2693(01)00524-X} {\bibfield  {journal} {\bibinfo  {journal} {Phys. Lett. B}\ }\textbf {\bibinfo {volume} {510}},\ \bibinfo {pages} {133} (\bibinfo {year} {2001}{\natexlab{b}})},\ \Eprint {https://arxiv.org/abs/hep-ph/0102009} {arXiv:hep-ph/0102009} \BibitemShut {NoStop}%
\bibitem [{\citenamefont {Ferreiro}\ \emph {et~al.}(2002)\citenamefont {Ferreiro}, \citenamefont {Iancu}, \citenamefont {Leonidov},\ and\ \citenamefont {McLerran}}]{Ferreiro:2001qy}%
  \BibitemOpen
  \bibfield  {author} {\bibinfo {author} {\bibfnamefont {E.}~\bibnamefont {Ferreiro}}, \bibinfo {author} {\bibfnamefont {E.}~\bibnamefont {Iancu}}, \bibinfo {author} {\bibfnamefont {A.}~\bibnamefont {Leonidov}},\ and\ \bibinfo {author} {\bibfnamefont {L.}~\bibnamefont {McLerran}},\ }\bibfield  {title} {\bibinfo {title} {{Nonlinear gluon evolution in the color glass condensate. 2.}},\ }\href {https://doi.org/10.1016/S0375-9474(01)01329-X} {\bibfield  {journal} {\bibinfo  {journal} {Nucl. Phys. A}\ }\textbf {\bibinfo {volume} {703}},\ \bibinfo {pages} {489} (\bibinfo {year} {2002})},\ \Eprint {https://arxiv.org/abs/hep-ph/0109115} {arXiv:hep-ph/0109115} \BibitemShut {NoStop}%
\bibitem [{\citenamefont {Balitsky}(1996)}]{Balitsky:1995ub}%
  \BibitemOpen
  \bibfield  {author} {\bibinfo {author} {\bibfnamefont {I.}~\bibnamefont {Balitsky}},\ }\bibfield  {title} {\bibinfo {title} {{Operator expansion for high-energy scattering}},\ }\href {https://doi.org/10.1016/0550-3213(95)00638-9} {\bibfield  {journal} {\bibinfo  {journal} {Nucl. Phys. B}\ }\textbf {\bibinfo {volume} {463}},\ \bibinfo {pages} {99} (\bibinfo {year} {1996})},\ \Eprint {https://arxiv.org/abs/hep-ph/9509348} {arXiv:hep-ph/9509348} \BibitemShut {NoStop}%
\bibitem [{\citenamefont {Balitsky}(1998)}]{Balitsky:1998kc}%
  \BibitemOpen
  \bibfield  {author} {\bibinfo {author} {\bibfnamefont {I.}~\bibnamefont {Balitsky}},\ }\bibfield  {title} {\bibinfo {title} {{Factorization for high-energy scattering}},\ }\href {https://doi.org/10.1103/PhysRevLett.81.2024} {\bibfield  {journal} {\bibinfo  {journal} {Phys. Rev. Lett.}\ }\textbf {\bibinfo {volume} {81}},\ \bibinfo {pages} {2024} (\bibinfo {year} {1998})},\ \Eprint {https://arxiv.org/abs/hep-ph/9807434} {arXiv:hep-ph/9807434} \BibitemShut {NoStop}%
\bibitem [{\citenamefont {Mueller}(1994)}]{Mueller:1993rr}%
  \BibitemOpen
  \bibfield  {author} {\bibinfo {author} {\bibfnamefont {A.~H.}\ \bibnamefont {Mueller}},\ }\bibfield  {title} {\bibinfo {title} {{Soft gluons in the infinite momentum wave function and the BFKL pomeron}},\ }\href {https://doi.org/10.1016/0550-3213(94)90116-3} {\bibfield  {journal} {\bibinfo  {journal} {Nucl. Phys. B}\ }\textbf {\bibinfo {volume} {415}},\ \bibinfo {pages} {373} (\bibinfo {year} {1994})}\BibitemShut {NoStop}%
\bibitem [{\citenamefont {McLerran}\ and\ \citenamefont {Venugopalan}(1994{\natexlab{a}})}]{McLerran:1993ni}%
  \BibitemOpen
  \bibfield  {author} {\bibinfo {author} {\bibfnamefont {L.~D.}\ \bibnamefont {McLerran}}\ and\ \bibinfo {author} {\bibfnamefont {R.}~\bibnamefont {Venugopalan}},\ }\bibfield  {title} {\bibinfo {title} {{Computing quark and gluon distribution functions for very large nuclei}},\ }\href {https://doi.org/10.1103/PhysRevD.49.2233} {\bibfield  {journal} {\bibinfo  {journal} {Phys. Rev. D}\ }\textbf {\bibinfo {volume} {49}},\ \bibinfo {pages} {2233} (\bibinfo {year} {1994}{\natexlab{a}})},\ \Eprint {https://arxiv.org/abs/hep-ph/9309289} {arXiv:hep-ph/9309289} \BibitemShut {NoStop}%
\bibitem [{\citenamefont {McLerran}\ and\ \citenamefont {Venugopalan}(1994{\natexlab{b}})}]{McLerran:1993ka}%
  \BibitemOpen
  \bibfield  {author} {\bibinfo {author} {\bibfnamefont {L.~D.}\ \bibnamefont {McLerran}}\ and\ \bibinfo {author} {\bibfnamefont {R.}~\bibnamefont {Venugopalan}},\ }\bibfield  {title} {\bibinfo {title} {{Gluon distribution functions for very large nuclei at small transverse momentum}},\ }\href {https://doi.org/10.1103/PhysRevD.49.3352} {\bibfield  {journal} {\bibinfo  {journal} {Phys. Rev. D}\ }\textbf {\bibinfo {volume} {49}},\ \bibinfo {pages} {3352} (\bibinfo {year} {1994}{\natexlab{b}})},\ \Eprint {https://arxiv.org/abs/hep-ph/9311205} {arXiv:hep-ph/9311205} \BibitemShut {NoStop}%
\bibitem [{\citenamefont {Gelis}\ \emph {et~al.}(2010)\citenamefont {Gelis}, \citenamefont {Iancu}, \citenamefont {Jalilian-Marian},\ and\ \citenamefont {Venugopalan}}]{Gelis:2010nm}%
  \BibitemOpen
  \bibfield  {author} {\bibinfo {author} {\bibfnamefont {F.}~\bibnamefont {Gelis}}, \bibinfo {author} {\bibfnamefont {E.}~\bibnamefont {Iancu}}, \bibinfo {author} {\bibfnamefont {J.}~\bibnamefont {Jalilian-Marian}},\ and\ \bibinfo {author} {\bibfnamefont {R.}~\bibnamefont {Venugopalan}},\ }\bibfield  {title} {\bibinfo {title} {{The Color Glass Condensate}},\ }\href {https://doi.org/10.1146/annurev.nucl.010909.083629} {\bibfield  {journal} {\bibinfo  {journal} {Ann. Rev. Nucl. Part. Sci.}\ }\textbf {\bibinfo {volume} {60}},\ \bibinfo {pages} {463} (\bibinfo {year} {2010})},\ \Eprint {https://arxiv.org/abs/1002.0333} {arXiv:1002.0333 [hep-ph]} \BibitemShut {NoStop}%
\bibitem [{\citenamefont {Shanahan}\ and\ \citenamefont {Detmold}(2019)}]{Shanahan:2018pib}%
  \BibitemOpen
  \bibfield  {author} {\bibinfo {author} {\bibfnamefont {P.~E.}\ \bibnamefont {Shanahan}}\ and\ \bibinfo {author} {\bibfnamefont {W.}~\bibnamefont {Detmold}},\ }\bibfield  {title} {\bibinfo {title} {{Gluon gravitational form factors of the proton and the pion from lattice QCD}},\ }\href {https://doi.org/10.1103/PhysRevLett.122.072003} {\bibfield  {journal} {\bibinfo  {journal} {Phys. Rev. Lett.}\ }\textbf {\bibinfo {volume} {122}},\ \bibinfo {pages} {072003} (\bibinfo {year} {2019})},\ \Eprint {https://arxiv.org/abs/1810.04626} {arXiv:1810.04626} \BibitemShut {NoStop}%
\bibitem [{\citenamefont {Hackett}\ \emph {et~al.}(2024)\citenamefont {Hackett}, \citenamefont {Pefkou},\ and\ \citenamefont {Shanahan}}]{Hackett:2023rif}%
  \BibitemOpen
  \bibfield  {author} {\bibinfo {author} {\bibfnamefont {D.~C.}\ \bibnamefont {Hackett}}, \bibinfo {author} {\bibfnamefont {D.~A.}\ \bibnamefont {Pefkou}},\ and\ \bibinfo {author} {\bibfnamefont {P.~E.}\ \bibnamefont {Shanahan}},\ }\bibfield  {title} {\bibinfo {title} {{Gravitational form factors of the proton from lattice QCD}},\ }\href {https://doi.org/10.1103/PhysRevLett.132.251904} {\bibfield  {journal} {\bibinfo  {journal} {Phys. Rev. Lett.}\ }\textbf {\bibinfo {volume} {132}},\ \bibinfo {pages} {251904} (\bibinfo {year} {2024})},\ \Eprint {https://arxiv.org/abs/2310.08484} {arXiv:2310.08484} \BibitemShut {NoStop}%
\bibitem [{\citenamefont {Hackett}\ \emph {et~al.}(2023)\citenamefont {Hackett}, \citenamefont {Oare}, \citenamefont {Pefkou},\ and\ \citenamefont {Shanahan}}]{Hackett:2023nkr}%
  \BibitemOpen
  \bibfield  {author} {\bibinfo {author} {\bibfnamefont {D.~C.}\ \bibnamefont {Hackett}}, \bibinfo {author} {\bibfnamefont {P.~R.}\ \bibnamefont {Oare}}, \bibinfo {author} {\bibfnamefont {D.~A.}\ \bibnamefont {Pefkou}},\ and\ \bibinfo {author} {\bibfnamefont {P.~E.}\ \bibnamefont {Shanahan}},\ }\bibfield  {title} {\bibinfo {title} {{Gravitational form factors of the pion from lattice QCD}},\ }\href {https://doi.org/10.1103/PhysRevD.108.114504} {\bibfield  {journal} {\bibinfo  {journal} {Phys. Rev. D}\ }\textbf {\bibinfo {volume} {108}},\ \bibinfo {pages} {114504} (\bibinfo {year} {2023})},\ \Eprint {https://arxiv.org/abs/2307.11707} {arXiv:2307.11707} \BibitemShut {NoStop}%
\bibitem [{\citenamefont {Wang}\ \emph {et~al.}(2024)\citenamefont {Wang} \emph {et~al.}}]{Wang:2024lrm}%
  \BibitemOpen
  \bibfield  {author} {\bibinfo {author} {\bibfnamefont {B.}~\bibnamefont {Wang}} \emph {et~al.} (\bibinfo {collaboration} {$\chi$QCD}),\ }\bibfield  {title} {\bibinfo {title} {{Trace anomaly form factors from lattice QCD}},\ }\href {https://doi.org/10.1103/PhysRevD.109.094504} {\bibfield  {journal} {\bibinfo  {journal} {Phys. Rev. D}\ }\textbf {\bibinfo {volume} {109}},\ \bibinfo {pages} {094504} (\bibinfo {year} {2024})},\ \Eprint {https://arxiv.org/abs/2401.05496} {arXiv:2401.05496} \BibitemShut {NoStop}%
\bibitem [{\citenamefont {Burkert}\ \emph {et~al.}(2018)\citenamefont {Burkert}, \citenamefont {Elouadrhiri},\ and\ \citenamefont {Girod}}]{Burkert:2018bqq}%
  \BibitemOpen
  \bibfield  {author} {\bibinfo {author} {\bibfnamefont {V.~D.}\ \bibnamefont {Burkert}}, \bibinfo {author} {\bibfnamefont {L.}~\bibnamefont {Elouadrhiri}},\ and\ \bibinfo {author} {\bibfnamefont {F.~X.}\ \bibnamefont {Girod}},\ }\bibfield  {title} {\bibinfo {title} {{The pressure distribution inside the proton}},\ }\href {https://doi.org/10.1038/s41586-018-0060-z} {\bibfield  {journal} {\bibinfo  {journal} {Nature}\ }\textbf {\bibinfo {volume} {557}},\ \bibinfo {pages} {396} (\bibinfo {year} {2018})}\BibitemShut {NoStop}%
\bibitem [{\citenamefont {Diehl}\ and\ \citenamefont {Ivanov}(2007)}]{Diehl:2007jb}%
  \BibitemOpen
  \bibfield  {author} {\bibinfo {author} {\bibfnamefont {M.}~\bibnamefont {Diehl}}\ and\ \bibinfo {author} {\bibfnamefont {D.~Y.}\ \bibnamefont {Ivanov}},\ }\bibfield  {title} {\bibinfo {title} {{Dispersion representations for hard exclusive processes: beyond the Born approximation}},\ }\href {https://doi.org/10.1140/epjc/s10052-007-0401-9} {\bibfield  {journal} {\bibinfo  {journal} {Eur. Phys. J. C}\ }\textbf {\bibinfo {volume} {52}},\ \bibinfo {pages} {919} (\bibinfo {year} {2007})},\ \Eprint {https://arxiv.org/abs/0707.0351} {arXiv:0707.0351 [hep-ph]} \BibitemShut {NoStop}%
\bibitem [{\citenamefont {Anikin}\ and\ \citenamefont {Teryaev}(2008)}]{Anikin:2007tx}%
  \BibitemOpen
  \bibfield  {author} {\bibinfo {author} {\bibfnamefont {I.~V.}\ \bibnamefont {Anikin}}\ and\ \bibinfo {author} {\bibfnamefont {O.~V.}\ \bibnamefont {Teryaev}},\ }\bibfield  {title} {\bibinfo {title} {{Dispersion relations and QCD factorization in hard reactions}},\ }\href@noop {} {\bibfield  {journal} {\bibinfo  {journal} {Fizika B}\ }\textbf {\bibinfo {volume} {17}},\ \bibinfo {pages} {151} (\bibinfo {year} {2008})},\ \Eprint {https://arxiv.org/abs/0710.4211} {arXiv:0710.4211 [hep-ph]} \BibitemShut {NoStop}%
\bibitem [{\citenamefont {Pasquini}\ \emph {et~al.}(2014)\citenamefont {Pasquini}, \citenamefont {Polyakov},\ and\ \citenamefont {Vanderhaeghen}}]{Pasquini:2014vua}%
  \BibitemOpen
  \bibfield  {author} {\bibinfo {author} {\bibfnamefont {B.}~\bibnamefont {Pasquini}}, \bibinfo {author} {\bibfnamefont {M.~V.}\ \bibnamefont {Polyakov}},\ and\ \bibinfo {author} {\bibfnamefont {M.}~\bibnamefont {Vanderhaeghen}},\ }\bibfield  {title} {\bibinfo {title} {{Dispersive evaluation of the D-term form factor in deeply virtual Compton scattering}},\ }\href {https://doi.org/10.1016/j.physletb.2014.10.047} {\bibfield  {journal} {\bibinfo  {journal} {Phys. Lett. B}\ }\textbf {\bibinfo {volume} {739}},\ \bibinfo {pages} {133} (\bibinfo {year} {2014})},\ \Eprint {https://arxiv.org/abs/1407.5960} {arXiv:1407.5960 [hep-ph]} \BibitemShut {NoStop}%
\bibitem [{\citenamefont {Duran}\ \emph {et~al.}(2023)\citenamefont {Duran} \emph {et~al.}}]{Duran:2022xag}%
  \BibitemOpen
  \bibfield  {author} {\bibinfo {author} {\bibfnamefont {B.}~\bibnamefont {Duran}} \emph {et~al.},\ }\bibfield  {title} {\bibinfo {title} {{Determining the gluonic gravitational form factors of the proton}},\ }\href {https://doi.org/10.1038/s41586-023-05730-4} {\bibfield  {journal} {\bibinfo  {journal} {Nature}\ }\textbf {\bibinfo {volume} {615}},\ \bibinfo {pages} {813} (\bibinfo {year} {2023})},\ \Eprint {https://arxiv.org/abs/2207.05212} {arXiv:2207.05212 [nucl-ex]} \BibitemShut {NoStop}%
\bibitem [{\citenamefont {Mamo}\ and\ \citenamefont {Zahed}(2020)}]{Mamo:2019mka}%
  \BibitemOpen
  \bibfield  {author} {\bibinfo {author} {\bibfnamefont {K.~A.}\ \bibnamefont {Mamo}}\ and\ \bibinfo {author} {\bibfnamefont {I.}~\bibnamefont {Zahed}},\ }\bibfield  {title} {\bibinfo {title} {{Diffractive photoproduction of $J/\psi$ and $\Upsilon$ using holographic QCD: gravitational form factors and GPD of gluons in the proton}},\ }\href {https://doi.org/10.1103/PhysRevD.101.086003} {\bibfield  {journal} {\bibinfo  {journal} {Phys. Rev. D}\ }\textbf {\bibinfo {volume} {101}},\ \bibinfo {pages} {086003} (\bibinfo {year} {2020})},\ \Eprint {https://arxiv.org/abs/1910.04707} {arXiv:1910.04707 [hep-ph]} \BibitemShut {NoStop}%
\bibitem [{\citenamefont {Mamo}\ and\ \citenamefont {Zahed}(2022)}]{Mamo:2022eui}%
  \BibitemOpen
  \bibfield  {author} {\bibinfo {author} {\bibfnamefont {K.~A.}\ \bibnamefont {Mamo}}\ and\ \bibinfo {author} {\bibfnamefont {I.}~\bibnamefont {Zahed}},\ }\bibfield  {title} {\bibinfo {title} {{J/{\ensuremath{\psi}} near threshold in holographic QCD: A and D gravitational form factors}},\ }\href {https://doi.org/10.1103/PhysRevD.106.086004} {\bibfield  {journal} {\bibinfo  {journal} {Phys. Rev. D}\ }\textbf {\bibinfo {volume} {106}},\ \bibinfo {pages} {086004} (\bibinfo {year} {2022})},\ \Eprint {https://arxiv.org/abs/2204.08857} {arXiv:2204.08857 [hep-ph]} \BibitemShut {NoStop}%
\bibitem [{\citenamefont {Sun}\ \emph {et~al.}(2021)\citenamefont {Sun}, \citenamefont {Tong},\ and\ \citenamefont {Yuan}}]{Sun:2021gmi}%
  \BibitemOpen
  \bibfield  {author} {\bibinfo {author} {\bibfnamefont {P.}~\bibnamefont {Sun}}, \bibinfo {author} {\bibfnamefont {X.-B.}\ \bibnamefont {Tong}},\ and\ \bibinfo {author} {\bibfnamefont {F.}~\bibnamefont {Yuan}},\ }\bibfield  {title} {\bibinfo {title} {{Perturbative QCD analysis of near threshold heavy quarkonium photoproduction at large momentum transfer}},\ }\href {https://doi.org/10.1016/j.physletb.2021.136655} {\bibfield  {journal} {\bibinfo  {journal} {Phys. Lett. B}\ }\textbf {\bibinfo {volume} {822}},\ \bibinfo {pages} {136655} (\bibinfo {year} {2021})},\ \Eprint {https://arxiv.org/abs/2103.12047} {arXiv:2103.12047 [hep-ph]} \BibitemShut {NoStop}%
\bibitem [{\citenamefont {Chernodub}\ \emph {et~al.}(2023)\citenamefont {Chernodub}, \citenamefont {Goy}, \citenamefont {Molochkov},\ and\ \citenamefont {Tanashkin}}]{Chernodub:2023dok}%
  \BibitemOpen
  \bibfield  {author} {\bibinfo {author} {\bibfnamefont {M.~N.}\ \bibnamefont {Chernodub}}, \bibinfo {author} {\bibfnamefont {V.~A.}\ \bibnamefont {Goy}}, \bibinfo {author} {\bibfnamefont {A.~V.}\ \bibnamefont {Molochkov}},\ and\ \bibinfo {author} {\bibfnamefont {A.~S.}\ \bibnamefont {Tanashkin}},\ }\bibfield  {title} {\bibinfo {title} {{Boundary states and non-Abelian Casimir effect in lattice Yang-Mills theory}},\ }\href {https://doi.org/10.1103/PhysRevD.108.014515} {\bibfield  {journal} {\bibinfo  {journal} {Phys. Rev. D}\ }\textbf {\bibinfo {volume} {108}},\ \bibinfo {pages} {014515} (\bibinfo {year} {2023})},\ \Eprint {https://arxiv.org/abs/2302.00376} {arXiv:2302.00376 [hep-lat]} \BibitemShut {NoStop}%
\bibitem [{\citenamefont {Polyakov}(2003)}]{Polyakov:2002yz}%
  \BibitemOpen
  \bibfield  {author} {\bibinfo {author} {\bibfnamefont {M.~V.}\ \bibnamefont {Polyakov}},\ }\bibfield  {title} {\bibinfo {title} {{Generalized parton distributions and strong forces inside nucleons and nuclei}},\ }\href {https://doi.org/10.1016/S0370-2693(03)00036-4} {\bibfield  {journal} {\bibinfo  {journal} {Phys. Lett. B}\ }\textbf {\bibinfo {volume} {555}},\ \bibinfo {pages} {57} (\bibinfo {year} {2003})},\ \Eprint {https://arxiv.org/abs/hep-ph/0210165} {arXiv:hep-ph/0210165} \BibitemShut {NoStop}%
\bibitem [{\citenamefont {Mamo}\ and\ \citenamefont {Zahed}(2021)}]{Mamo:2021krl}%
  \BibitemOpen
  \bibfield  {author} {\bibinfo {author} {\bibfnamefont {K.~A.}\ \bibnamefont {Mamo}}\ and\ \bibinfo {author} {\bibfnamefont {I.}~\bibnamefont {Zahed}},\ }\bibfield  {title} {\bibinfo {title} {{Nucleon mass radii and distribution: Holographic QCD, Lattice QCD and GlueX data}},\ }\href {https://doi.org/10.1103/PhysRevD.103.094010} {\bibfield  {journal} {\bibinfo  {journal} {Phys. Rev. D}\ }\textbf {\bibinfo {volume} {103}},\ \bibinfo {pages} {094010} (\bibinfo {year} {2021})},\ \Eprint {https://arxiv.org/abs/2103.03186} {arXiv:2103.03186} \BibitemShut {NoStop}%
\bibitem [{\citenamefont {Cao}\ \emph {et~al.}(2025)\citenamefont {Cao}, \citenamefont {Guo}, \citenamefont {Li},\ and\ \citenamefont {Yao}}]{cao2025dispersivedeterminationnucleongravitational}%
  \BibitemOpen
  \bibfield  {author} {\bibinfo {author} {\bibfnamefont {X.-H.}\ \bibnamefont {Cao}}, \bibinfo {author} {\bibfnamefont {F.-K.}\ \bibnamefont {Guo}}, \bibinfo {author} {\bibfnamefont {Q.-Z.}\ \bibnamefont {Li}},\ and\ \bibinfo {author} {\bibfnamefont {D.-L.}\ \bibnamefont {Yao}},\ }\href {https://arxiv.org/abs/2411.13398} {\bibinfo {title} {Dispersive determination of nucleon gravitational form factors}} (\bibinfo {year} {2025}),\ \Eprint {https://arxiv.org/abs/2411.13398} {arXiv:2411.13398 [hep-ph]} \BibitemShut {NoStop}%
\bibitem [{\citenamefont {Broniowski}\ and\ \citenamefont {Ruiz~Arriola}(2024)}]{Broniowski:2024oyk}%
  \BibitemOpen
  \bibfield  {author} {\bibinfo {author} {\bibfnamefont {W.}~\bibnamefont {Broniowski}}\ and\ \bibinfo {author} {\bibfnamefont {E.}~\bibnamefont {Ruiz~Arriola}},\ }\bibfield  {title} {\bibinfo {title} {{Gravitational form factors of the pion and meson dominance}},\ }\href {https://doi.org/10.1016/j.physletb.2024.139138} {\bibfield  {journal} {\bibinfo  {journal} {Phys. Lett. B}\ }\textbf {\bibinfo {volume} {859}},\ \bibinfo {pages} {139138} (\bibinfo {year} {2024})},\ \Eprint {https://arxiv.org/abs/2405.07815} {arXiv:2405.07815 [hep-ph]} \BibitemShut {NoStop}%
\bibitem [{\citenamefont {Broniowski}\ and\ \citenamefont {Arriola}(2025)}]{broniowski2025gravitationalformfactorsmechanical}%
  \BibitemOpen
  \bibfield  {author} {\bibinfo {author} {\bibfnamefont {W.}~\bibnamefont {Broniowski}}\ and\ \bibinfo {author} {\bibfnamefont {E.~R.}\ \bibnamefont {Arriola}},\ }\href {https://arxiv.org/abs/2503.09297} {\bibinfo {title} {Gravitational form factors and mechanical properties of the nucleon in a meson dominance approach}} (\bibinfo {year} {2025}),\ \Eprint {https://arxiv.org/abs/2503.09297} {arXiv:2503.09297 [hep-ph]} \BibitemShut {NoStop}%
\bibitem [{\citenamefont {Lorc{\'e}}\ and\ \citenamefont {Schweitzer}(2025)}]{Lorce:2025oot}%
  \BibitemOpen
  \bibfield  {author} {\bibinfo {author} {\bibfnamefont {C.}~\bibnamefont {Lorc{\'e}}}\ and\ \bibinfo {author} {\bibfnamefont {P.}~\bibnamefont {Schweitzer}},\ }\bibfield  {title} {\bibinfo {title} {{Pressure inside hadrons: criticism, conjectures, and all that}},\ }\href {https://doi.org/10.5506/APhysPolB.56.3-A17} {\bibfield  {journal} {\bibinfo  {journal} {Acta Phys. Polon. B}\ }\textbf {\bibinfo {volume} {56}},\ \bibinfo {pages} {3} (\bibinfo {year} {2025})},\ \Eprint {https://arxiv.org/abs/2501.04622} {arXiv:2501.04622 [hep-ph]} \BibitemShut {NoStop}%
\bibitem [{\citenamefont {Liu}\ \emph {et~al.}(2023)\citenamefont {Liu}, \citenamefont {Nowak},\ and\ \citenamefont {Zahed}}]{Liu:2023zno}%
  \BibitemOpen
  \bibfield  {author} {\bibinfo {author} {\bibfnamefont {Y.}~\bibnamefont {Liu}}, \bibinfo {author} {\bibfnamefont {M.~A.}\ \bibnamefont {Nowak}},\ and\ \bibinfo {author} {\bibfnamefont {I.}~\bibnamefont {Zahed}},\ }\bibfield  {title} {\bibinfo {title} {{Nambu–Goto string in QCD: dipole interactions, scattering, and entanglement}},\ }\href {https://doi.org/10.1103/PhysRevD.108.094025} {\bibfield  {journal} {\bibinfo  {journal} {Phys. Rev. D}\ }\textbf {\bibinfo {volume} {108}},\ \bibinfo {pages} {094025} (\bibinfo {year} {2023})},\ \Eprint {https://arxiv.org/abs/2301.06154} {arXiv:2301.06154} \BibitemShut {NoStop}%
\bibitem [{\citenamefont {Low}\ and\ \citenamefont {Yin}(2025)}]{Low:2024hvn}%
  \BibitemOpen
  \bibfield  {author} {\bibinfo {author} {\bibfnamefont {I.}~\bibnamefont {Low}}\ and\ \bibinfo {author} {\bibfnamefont {Z.}~\bibnamefont {Yin}},\ }\bibfield  {title} {\bibinfo {title} {{Elastic cross section is entanglement entropy}},\ }\href {https://doi.org/10.1103/PhysRevD.111.065027} {\bibfield  {journal} {\bibinfo  {journal} {Phys. Rev. D}\ }\textbf {\bibinfo {volume} {111}},\ \bibinfo {pages} {065027} (\bibinfo {year} {2025})},\ \Eprint {https://arxiv.org/abs/2410.22414} {arXiv:2410.22414 [hep-th]} \BibitemShut {NoStop}%
\bibitem [{\citenamefont {Dosch}\ \emph {et~al.}(2024)\citenamefont {Dosch}, \citenamefont {de~Teramond},\ and\ \citenamefont {Brodsky}}]{Dosch:2023bxj}%
  \BibitemOpen
  \bibfield  {author} {\bibinfo {author} {\bibfnamefont {H.~G.}\ \bibnamefont {Dosch}}, \bibinfo {author} {\bibfnamefont {G.~F.}\ \bibnamefont {de~Teramond}},\ and\ \bibinfo {author} {\bibfnamefont {S.~J.}\ \bibnamefont {Brodsky}},\ }\bibfield  {title} {\bibinfo {title} {{Entropy from entangled parton states and high-energy scattering behavior}},\ }\href {https://doi.org/10.1016/j.physletb.2024.138521} {\bibfield  {journal} {\bibinfo  {journal} {Phys. Lett. B}\ }\textbf {\bibinfo {volume} {850}},\ \bibinfo {pages} {138521} (\bibinfo {year} {2024})},\ \Eprint {https://arxiv.org/abs/2304.14207} {arXiv:2304.14207 [hep-ph]} \BibitemShut {NoStop}%
\bibitem [{\citenamefont {Kovner}\ and\ \citenamefont {Lublinsky}(2015)}]{Kovner:2015hga}%
  \BibitemOpen
  \bibfield  {author} {\bibinfo {author} {\bibfnamefont {A.}~\bibnamefont {Kovner}}\ and\ \bibinfo {author} {\bibfnamefont {M.}~\bibnamefont {Lublinsky}},\ }\bibfield  {title} {\bibinfo {title} {{Entanglement entropy and entropy production in the Color Glass Condensate framework}},\ }\href {https://doi.org/10.1103/PhysRevD.92.034016} {\bibfield  {journal} {\bibinfo  {journal} {Phys. Rev. D}\ }\textbf {\bibinfo {volume} {92}},\ \bibinfo {pages} {034016} (\bibinfo {year} {2015})},\ \Eprint {https://arxiv.org/abs/1506.05394} {arXiv:1506.05394 [hep-ph]} \BibitemShut {NoStop}%
\bibitem [{\citenamefont {Kovner}\ \emph {et~al.}(2019)\citenamefont {Kovner}, \citenamefont {Lublinsky},\ and\ \citenamefont {Serino}}]{Kovner:2018rbf}%
  \BibitemOpen
  \bibfield  {author} {\bibinfo {author} {\bibfnamefont {A.}~\bibnamefont {Kovner}}, \bibinfo {author} {\bibfnamefont {M.}~\bibnamefont {Lublinsky}},\ and\ \bibinfo {author} {\bibfnamefont {M.}~\bibnamefont {Serino}},\ }\bibfield  {title} {\bibinfo {title} {{Entanglement entropy, entropy production and time evolution in high energy QCD}},\ }\href {https://doi.org/10.1016/j.physletb.2018.10.043} {\bibfield  {journal} {\bibinfo  {journal} {Phys. Lett. B}\ }\textbf {\bibinfo {volume} {792}},\ \bibinfo {pages} {4} (\bibinfo {year} {2019})},\ \Eprint {https://arxiv.org/abs/1806.01089} {arXiv:1806.01089 [hep-ph]} \BibitemShut {NoStop}%
\bibitem [{\citenamefont {Tu}\ \emph {et~al.}(2020)\citenamefont {Tu}, \citenamefont {Kharzeev},\ and\ \citenamefont {Ullrich}}]{Tu:2019ouv}%
  \BibitemOpen
  \bibfield  {author} {\bibinfo {author} {\bibfnamefont {Z.}~\bibnamefont {Tu}}, \bibinfo {author} {\bibfnamefont {D.~E.}\ \bibnamefont {Kharzeev}},\ and\ \bibinfo {author} {\bibfnamefont {T.}~\bibnamefont {Ullrich}},\ }\bibfield  {title} {\bibinfo {title} {{Einstein-Podolsky-Rosen Paradox and Quantum Entanglement at Subnucleonic Scales}},\ }\href {https://doi.org/10.1103/PhysRevLett.124.062001} {\bibfield  {journal} {\bibinfo  {journal} {Phys. Rev. Lett.}\ }\textbf {\bibinfo {volume} {124}},\ \bibinfo {pages} {062001} (\bibinfo {year} {2020})},\ \Eprint {https://arxiv.org/abs/1904.11974} {arXiv:1904.11974 [hep-ph]} \BibitemShut {NoStop}%
\bibitem [{\citenamefont {Ramos}\ and\ \citenamefont {Machado}(2020)}]{Ramos:2020kaj}%
  \BibitemOpen
  \bibfield  {author} {\bibinfo {author} {\bibfnamefont {G.~S.}\ \bibnamefont {Ramos}}\ and\ \bibinfo {author} {\bibfnamefont {M.~V.~T.}\ \bibnamefont {Machado}},\ }\bibfield  {title} {\bibinfo {title} {{Investigating entanglement entropy at small-x in DIS off protons and nuclei}},\ }\href {https://doi.org/10.1103/PhysRevD.101.074040} {\bibfield  {journal} {\bibinfo  {journal} {Phys. Rev. D}\ }\textbf {\bibinfo {volume} {101}},\ \bibinfo {pages} {074040} (\bibinfo {year} {2020})},\ \Eprint {https://arxiv.org/abs/2003.05008} {arXiv:2003.05008 [hep-ph]} \BibitemShut {NoStop}%
\bibitem [{\citenamefont {Gotsman}\ and\ \citenamefont {Levin}(2020)}]{Gotsman:2020bjc}%
  \BibitemOpen
  \bibfield  {author} {\bibinfo {author} {\bibfnamefont {E.}~\bibnamefont {Gotsman}}\ and\ \bibinfo {author} {\bibfnamefont {E.}~\bibnamefont {Levin}},\ }\bibfield  {title} {\bibinfo {title} {{High energy QCD: multiplicity distribution and entanglement entropy}},\ }\href {https://doi.org/10.1103/PhysRevD.102.074008} {\bibfield  {journal} {\bibinfo  {journal} {Phys. Rev. D}\ }\textbf {\bibinfo {volume} {102}},\ \bibinfo {pages} {074008} (\bibinfo {year} {2020})},\ \Eprint {https://arxiv.org/abs/2006.11793} {arXiv:2006.11793 [hep-ph]} \BibitemShut {NoStop}%
\bibitem [{\citenamefont {Andreev}\ \emph {et~al.}(2021)\citenamefont {Andreev} \emph {et~al.}}]{H1:2020zpd}%
  \BibitemOpen
  \bibfield  {author} {\bibinfo {author} {\bibfnamefont {V.}~\bibnamefont {Andreev}} \emph {et~al.} (\bibinfo {collaboration} {H1}),\ }\bibfield  {title} {\bibinfo {title} {{Measurement of charged particle multiplicity distributions in DIS at HERA and its implication to entanglement entropy of partons}},\ }\href {https://doi.org/10.1140/epjc/s10052-021-08896-1} {\bibfield  {journal} {\bibinfo  {journal} {Eur. Phys. J. C}\ }\textbf {\bibinfo {volume} {81}},\ \bibinfo {pages} {212} (\bibinfo {year} {2021})},\ \Eprint {https://arxiv.org/abs/2011.01812} {arXiv:2011.01812 [hep-ex]} \BibitemShut {NoStop}%
\bibitem [{\citenamefont {Kharzeev}\ and\ \citenamefont {Levin}(2021)}]{Kharzeev:2021yyf}%
  \BibitemOpen
  \bibfield  {author} {\bibinfo {author} {\bibfnamefont {D.~E.}\ \bibnamefont {Kharzeev}}\ and\ \bibinfo {author} {\bibfnamefont {E.}~\bibnamefont {Levin}},\ }\bibfield  {title} {\bibinfo {title} {{Deep inelastic scattering as a probe of entanglement: Confronting experimental data}},\ }\href {https://doi.org/10.1103/PhysRevD.104.L031503} {\bibfield  {journal} {\bibinfo  {journal} {Phys. Rev. D}\ }\textbf {\bibinfo {volume} {104}},\ \bibinfo {pages} {L031503} (\bibinfo {year} {2021})},\ \Eprint {https://arxiv.org/abs/2102.09773} {arXiv:2102.09773 [hep-ph]} \BibitemShut {NoStop}%
\bibitem [{\citenamefont {Hentschinski}\ and\ \citenamefont {Kutak}(2022)}]{Hentschinski:2021aux}%
  \BibitemOpen
  \bibfield  {author} {\bibinfo {author} {\bibfnamefont {M.}~\bibnamefont {Hentschinski}}\ and\ \bibinfo {author} {\bibfnamefont {K.}~\bibnamefont {Kutak}},\ }\bibfield  {title} {\bibinfo {title} {{Evidence for the maximally entangled low x proton in Deep Inelastic Scattering from H1 data}},\ }\href {https://doi.org/10.1140/epjc/s10052-022-10056-y} {\bibfield  {journal} {\bibinfo  {journal} {Eur. Phys. J. C}\ }\textbf {\bibinfo {volume} {82}},\ \bibinfo {pages} {111} (\bibinfo {year} {2022})},\ \bibinfo {note} {[Erratum: Eur.Phys.J.C 83, 1147 (2023)]},\ \Eprint {https://arxiv.org/abs/2110.06156} {arXiv:2110.06156 [hep-ph]} \BibitemShut {NoStop}%
\bibitem [{\citenamefont {Hentschinski}\ \emph {et~al.}(2022)\citenamefont {Hentschinski}, \citenamefont {Kutak},\ and\ \citenamefont {Straka}}]{Hentschinski:2022rsa}%
  \BibitemOpen
  \bibfield  {author} {\bibinfo {author} {\bibfnamefont {M.}~\bibnamefont {Hentschinski}}, \bibinfo {author} {\bibfnamefont {K.}~\bibnamefont {Kutak}},\ and\ \bibinfo {author} {\bibfnamefont {R.}~\bibnamefont {Straka}},\ }\bibfield  {title} {\bibinfo {title} {{Maximally entangled proton and charged hadron multiplicity in Deep Inelastic Scattering}},\ }\href {https://doi.org/10.1140/epjc/s10052-022-11122-1} {\bibfield  {journal} {\bibinfo  {journal} {Eur. Phys. J. C}\ }\textbf {\bibinfo {volume} {82}},\ \bibinfo {pages} {1147} (\bibinfo {year} {2022})},\ \Eprint {https://arxiv.org/abs/2207.09430} {arXiv:2207.09430 [hep-ph]} \BibitemShut {NoStop}%
\bibitem [{\citenamefont {Hentschinski}\ \emph {et~al.}(2023)\citenamefont {Hentschinski}, \citenamefont {Kharzeev}, \citenamefont {Kutak},\ and\ \citenamefont {Tu}}]{Hentschinski:2023izh}%
  \BibitemOpen
  \bibfield  {author} {\bibinfo {author} {\bibfnamefont {M.}~\bibnamefont {Hentschinski}}, \bibinfo {author} {\bibfnamefont {D.~E.}\ \bibnamefont {Kharzeev}}, \bibinfo {author} {\bibfnamefont {K.}~\bibnamefont {Kutak}},\ and\ \bibinfo {author} {\bibfnamefont {Z.}~\bibnamefont {Tu}},\ }\bibfield  {title} {\bibinfo {title} {{Probing the Onset of Maximal Entanglement inside the Proton in Diffractive Deep Inelastic Scattering}},\ }\href {https://doi.org/10.1103/PhysRevLett.131.241901} {\bibfield  {journal} {\bibinfo  {journal} {Phys. Rev. Lett.}\ }\textbf {\bibinfo {volume} {131}},\ \bibinfo {pages} {241901} (\bibinfo {year} {2023})},\ \Eprint {https://arxiv.org/abs/2305.03069} {arXiv:2305.03069 [hep-ph]} \BibitemShut {NoStop}%
\bibitem [{\citenamefont {Hentschinski}\ \emph {et~al.}(2024)\citenamefont {Hentschinski}, \citenamefont {Kharzeev}, \citenamefont {Kutak},\ and\ \citenamefont {Tu}}]{Hentschinski:2024gaa}%
  \BibitemOpen
  \bibfield  {author} {\bibinfo {author} {\bibfnamefont {M.}~\bibnamefont {Hentschinski}}, \bibinfo {author} {\bibfnamefont {D.~E.}\ \bibnamefont {Kharzeev}}, \bibinfo {author} {\bibfnamefont {K.}~\bibnamefont {Kutak}},\ and\ \bibinfo {author} {\bibfnamefont {Z.}~\bibnamefont {Tu}},\ }\bibfield  {title} {\bibinfo {title} {{QCD evolution of entanglement entropy}},\ }\href {https://doi.org/10.1088/1361-6633/ad910b} {\bibfield  {journal} {\bibinfo  {journal} {Rept. Prog. Phys.}\ }\textbf {\bibinfo {volume} {87}},\ \bibinfo {pages} {120501} (\bibinfo {year} {2024})},\ \Eprint {https://arxiv.org/abs/2408.01259} {arXiv:2408.01259 [hep-ph]} \BibitemShut {NoStop}%
\bibitem [{\citenamefont {Datta}\ \emph {et~al.}(2025)\citenamefont {Datta}, \citenamefont {Deshpande}, \citenamefont {Kharzeev}, \citenamefont {Na{\"\i}m},\ and\ \citenamefont {Tu}}]{Datta:2024hpn}%
  \BibitemOpen
  \bibfield  {author} {\bibinfo {author} {\bibfnamefont {J.}~\bibnamefont {Datta}}, \bibinfo {author} {\bibfnamefont {A.}~\bibnamefont {Deshpande}}, \bibinfo {author} {\bibfnamefont {D.~E.}\ \bibnamefont {Kharzeev}}, \bibinfo {author} {\bibfnamefont {C.~J.}\ \bibnamefont {Na{\"\i}m}},\ and\ \bibinfo {author} {\bibfnamefont {Z.}~\bibnamefont {Tu}},\ }\bibfield  {title} {\bibinfo {title} {{Entanglement as a Probe of Hadronization}},\ }\href {https://doi.org/10.1103/PhysRevLett.134.111902} {\bibfield  {journal} {\bibinfo  {journal} {Phys. Rev. Lett.}\ }\textbf {\bibinfo {volume} {134}},\ \bibinfo {pages} {111902} (\bibinfo {year} {2025})},\ \Eprint {https://arxiv.org/abs/2410.22331} {arXiv:2410.22331 [hep-ph]} \BibitemShut {NoStop}%
\bibitem [{\citenamefont {Moriggi}\ and\ \citenamefont {Machado}(2025)}]{Moriggi:2024tiz}%
  \BibitemOpen
  \bibfield  {author} {\bibinfo {author} {\bibfnamefont {L.~S.}\ \bibnamefont {Moriggi}}\ and\ \bibinfo {author} {\bibfnamefont {M.~V.~T.}\ \bibnamefont {Machado}},\ }\bibfield  {title} {\bibinfo {title} {{Precise determination of the Pomeron intercept via a scaling entropy analysis}},\ }\href {https://doi.org/10.1103/PhysRevD.111.014017} {\bibfield  {journal} {\bibinfo  {journal} {Phys. Rev. D}\ }\textbf {\bibinfo {volume} {111}},\ \bibinfo {pages} {014017} (\bibinfo {year} {2025})},\ \Eprint {https://arxiv.org/abs/2412.16348} {arXiv:2412.16348 [hep-ph]} \BibitemShut {NoStop}%
\bibitem [{\citenamefont {Pefkou}\ \emph {et~al.}(2022)\citenamefont {Pefkou}, \citenamefont {Hackett},\ and\ \citenamefont {Shanahan}}]{Pefkou:2021fni}%
  \BibitemOpen
  \bibfield  {author} {\bibinfo {author} {\bibfnamefont {D.~A.}\ \bibnamefont {Pefkou}}, \bibinfo {author} {\bibfnamefont {D.~C.}\ \bibnamefont {Hackett}},\ and\ \bibinfo {author} {\bibfnamefont {P.~E.}\ \bibnamefont {Shanahan}},\ }\bibfield  {title} {\bibinfo {title} {{Gluon gravitational structure of hadrons of different spin}},\ }\href {https://doi.org/10.1103/PhysRevD.105.054509} {\bibfield  {journal} {\bibinfo  {journal} {Phys. Rev. D}\ }\textbf {\bibinfo {volume} {105}},\ \bibinfo {pages} {054509} (\bibinfo {year} {2022})},\ \Eprint {https://arxiv.org/abs/2107.10368} {arXiv:2107.10368 [hep-lat]} \BibitemShut {NoStop}%
\bibitem [{\citenamefont {Brodsky}\ and\ \citenamefont {Farrar}(1973)}]{Brodsky:1973kr}%
  \BibitemOpen
  \bibfield  {author} {\bibinfo {author} {\bibfnamefont {S.~J.}\ \bibnamefont {Brodsky}}\ and\ \bibinfo {author} {\bibfnamefont {G.~R.}\ \bibnamefont {Farrar}},\ }\bibfield  {title} {\bibinfo {title} {{Scaling Laws at Large Transverse Momentum}},\ }\href {https://doi.org/10.1103/PhysRevLett.31.1153} {\bibfield  {journal} {\bibinfo  {journal} {Phys. Rev. Lett.}\ }\textbf {\bibinfo {volume} {31}},\ \bibinfo {pages} {1153} (\bibinfo {year} {1973})}\BibitemShut {NoStop}%
\bibitem [{\citenamefont {Brodsky}\ and\ \citenamefont {Chertok}(1976)}]{Brodsky:1976rz}%
  \BibitemOpen
  \bibfield  {author} {\bibinfo {author} {\bibfnamefont {S.~J.}\ \bibnamefont {Brodsky}}\ and\ \bibinfo {author} {\bibfnamefont {B.~T.}\ \bibnamefont {Chertok}},\ }\bibfield  {title} {\bibinfo {title} {{The Asymptotic Form-Factors of Hadrons and Nuclei and the Continuity of Particle and Nuclear Dynamics}},\ }\href {https://doi.org/10.1103/PhysRevD.14.3003} {\bibfield  {journal} {\bibinfo  {journal} {Phys. Rev. D}\ }\textbf {\bibinfo {volume} {14}},\ \bibinfo {pages} {3003} (\bibinfo {year} {1976})}\BibitemShut {NoStop}%
\bibitem [{\citenamefont {Brodsky}\ and\ \citenamefont {Lepage}(1981)}]{Brodsky:1980th}%
  \BibitemOpen
  \bibfield  {author} {\bibinfo {author} {\bibfnamefont {S.~J.}\ \bibnamefont {Brodsky}}\ and\ \bibinfo {author} {\bibfnamefont {G.~P.}\ \bibnamefont {Lepage}},\ }\bibfield  {title} {\bibinfo {title} {{The Synthesis of Quantum Chromodynamics and Nuclear Physics}},\ }\href {https://doi.org/10.1016/0375-9474(81)90712-0} {\bibfield  {journal} {\bibinfo  {journal} {Nucl. Phys. A}\ }\textbf {\bibinfo {volume} {353}},\ \bibinfo {pages} {247C} (\bibinfo {year} {1981})}\BibitemShut {NoStop}%
\bibitem [{\citenamefont {Brodsky}\ and\ \citenamefont {Hiller}(1983)}]{Brodsky:1983kb}%
  \BibitemOpen
  \bibfield  {author} {\bibinfo {author} {\bibfnamefont {S.~J.}\ \bibnamefont {Brodsky}}\ and\ \bibinfo {author} {\bibfnamefont {J.~R.}\ \bibnamefont {Hiller}},\ }\bibfield  {title} {\bibinfo {title} {{Reduced Nuclear Amplitudes in Quantum Chromodynamics}},\ }\href {https://doi.org/10.1103/PhysRevC.28.475} {\bibfield  {journal} {\bibinfo  {journal} {Phys. Rev. C}\ }\textbf {\bibinfo {volume} {28}},\ \bibinfo {pages} {475} (\bibinfo {year} {1983})}\BibitemShut {NoStop}%
\bibitem [{\citenamefont {Tong}\ \emph {et~al.}(2021)\citenamefont {Tong}, \citenamefont {Ma},\ and\ \citenamefont {Yuan}}]{Tong:2021ctu}%
  \BibitemOpen
  \bibfield  {author} {\bibinfo {author} {\bibfnamefont {X.}~\bibnamefont {Tong}}, \bibinfo {author} {\bibfnamefont {J.-P.}\ \bibnamefont {Ma}},\ and\ \bibinfo {author} {\bibfnamefont {F.}~\bibnamefont {Yuan}},\ }\bibfield  {title} {\bibinfo {title} {{Gluon gravitational form factors at large momentum transfer}},\ }\href {https://doi.org/10.1016/j.physletb.2021.136751} {\bibfield  {journal} {\bibinfo  {journal} {Phys. Lett. B}\ }\textbf {\bibinfo {volume} {823}},\ \bibinfo {pages} {136751} (\bibinfo {year} {2021})},\ \Eprint {https://arxiv.org/abs/2101.02395} {arXiv:2101.02395} \BibitemShut {NoStop}%
\bibitem [{\citenamefont {Tong}\ \emph {et~al.}(2022)\citenamefont {Tong}, \citenamefont {Ma},\ and\ \citenamefont {Yuan}}]{Tong:2022zax}%
  \BibitemOpen
  \bibfield  {author} {\bibinfo {author} {\bibfnamefont {X.}~\bibnamefont {Tong}}, \bibinfo {author} {\bibfnamefont {J.-P.}\ \bibnamefont {Ma}},\ and\ \bibinfo {author} {\bibfnamefont {F.}~\bibnamefont {Yuan}},\ }\bibfield  {title} {\bibinfo {title} {{Perturbative calculations of gravitational form factors at large momentum transfer}},\ }\href {https://doi.org/10.1007/JHEP10(2022)046} {\bibfield  {journal} {\bibinfo  {journal} {JHEP}\ }\textbf {\bibinfo {volume} {10}},\ \bibinfo {pages} {046}},\ \Eprint {https://arxiv.org/abs/2203.13493} {arXiv:2203.13493} \BibitemShut {NoStop}%
\bibitem [{\citenamefont {Xiong}\ and\ \citenamefont {Others}(2019)}]{Xiong:2019umf}%
  \BibitemOpen
  \bibfield  {author} {\bibinfo {author} {\bibfnamefont {W.}~\bibnamefont {Xiong}}\ and\ \bibinfo {author} {\bibnamefont {Others}} (\bibinfo {collaboration} {PRad}),\ }\bibfield  {title} {\bibinfo {title} {{A small proton charge radius from an electron\textendash proton scattering experiment}},\ }\href {https://doi.org/10.1038/s41586-019-1721-2} {\bibfield  {journal} {\bibinfo  {journal} {Nature}\ }\textbf {\bibinfo {volume} {575}},\ \bibinfo {pages} {147} (\bibinfo {year} {2019})}\BibitemShut {NoStop}%
\bibitem [{Par(2024)}]{ParticleDataGroup:2024cfk}%
  \BibitemOpen
  \bibfield  {title} {\bibinfo {title} {{Review of Particle Physics}},\ }\href {https://doi.org/10.1103/PhysRevD.110.030001} {\bibfield  {journal} {\bibinfo  {journal} {Phys. Rev. D}\ }\textbf {\bibinfo {volume} {110}},\ \bibinfo {pages} {030001} (\bibinfo {year} {2024})}\BibitemShut {NoStop}%
\bibitem [{\citenamefont {Angeli}\ and\ \citenamefont {Marinova}(2013)}]{Angeli:2013epw}%
  \BibitemOpen
  \bibfield  {author} {\bibinfo {author} {\bibfnamefont {I.}~\bibnamefont {Angeli}}\ and\ \bibinfo {author} {\bibfnamefont {K.~P.}\ \bibnamefont {Marinova}},\ }\bibfield  {title} {\bibinfo {title} {{Table of experimental nuclear ground state charge radii: An update}},\ }\href {https://doi.org/10.1016/j.adt.2011.12.006} {\bibfield  {journal} {\bibinfo  {journal} {Atom. Data Nucl. Data Tables}\ }\textbf {\bibinfo {volume} {99}},\ \bibinfo {pages} {69} (\bibinfo {year} {2013})}\BibitemShut {NoStop}%
\bibitem [{\citenamefont {Egloff}\ \emph {et~al.}(1979)\citenamefont {Egloff} \emph {et~al.}}]{Egloff:1979mg}%
  \BibitemOpen
  \bibfield  {author} {\bibinfo {author} {\bibfnamefont {R.~M.}\ \bibnamefont {Egloff}} \emph {et~al.},\ }\bibfield  {title} {\bibinfo {title} {{Measurements of Elastic Rho and Phi Meson Photoproduction Cross-Sections on Protons from 30 GeV to 180 GeV}},\ }\href {https://doi.org/10.1103/PhysRevLett.43.657} {\bibfield  {journal} {\bibinfo  {journal} {Phys. Rev. Lett.}\ }\textbf {\bibinfo {volume} {43}},\ \bibinfo {pages} {657} (\bibinfo {year} {1979})}\BibitemShut {NoStop}%
\bibitem [{\citenamefont {Busenitz}\ \emph {et~al.}(1989)\citenamefont {Busenitz} \emph {et~al.}}]{Busenitz:1989gq}%
  \BibitemOpen
  \bibfield  {author} {\bibinfo {author} {\bibfnamefont {J.}~\bibnamefont {Busenitz}} \emph {et~al.},\ }\bibfield  {title} {\bibinfo {title} {{High-energy Photoproduction of $\pi^+ \pi^- \pi^0$, $K^+ K^-$, and $P \bar{P}$ States}},\ }\href {https://doi.org/10.1103/PhysRevD.40.1} {\bibfield  {journal} {\bibinfo  {journal} {Phys. Rev. D}\ }\textbf {\bibinfo {volume} {40}},\ \bibinfo {pages} {1} (\bibinfo {year} {1989})}\BibitemShut {NoStop}%
\bibitem [{\citenamefont {Antchev}\ \emph {et~al.}(2019)\citenamefont {Antchev} \emph {et~al.}}]{TOTEM:2018hki}%
  \BibitemOpen
  \bibfield  {author} {\bibinfo {author} {\bibfnamefont {G.}~\bibnamefont {Antchev}} \emph {et~al.} (\bibinfo {collaboration} {TOTEM}),\ }\bibfield  {title} {\bibinfo {title} {{Elastic differential cross-section measurement at $\sqrt{s}=13$ TeV by TOTEM}},\ }\href {https://doi.org/10.1140/epjc/s10052-019-7346-7} {\bibfield  {journal} {\bibinfo  {journal} {Eur. Phys. J. C}\ }\textbf {\bibinfo {volume} {79}},\ \bibinfo {pages} {861} (\bibinfo {year} {2019})},\ \Eprint {https://arxiv.org/abs/1812.08283} {arXiv:1812.08283 [hep-ex]} \BibitemShut {NoStop}%
\bibitem [{\citenamefont {Workman}\ \emph {et~al.}(2022)\citenamefont {Workman} \emph {et~al.}}]{ParticleDataGroup:2022pth}%
  \BibitemOpen
  \bibfield  {author} {\bibinfo {author} {\bibfnamefont {R.~L.}\ \bibnamefont {Workman}} \emph {et~al.} (\bibinfo {collaboration} {Particle Data Group}),\ }\bibfield  {title} {\bibinfo {title} {{Review of Particle Physics}},\ }\href {https://doi.org/10.1093/ptep/ptac097} {\bibfield  {journal} {\bibinfo  {journal} {PTEP}\ }\textbf {\bibinfo {volume} {2022}},\ \bibinfo {pages} {083C01} (\bibinfo {year} {2022})}\BibitemShut {NoStop}%
\bibitem [{\citenamefont {Aad}\ \emph {et~al.}(2023)\citenamefont {Aad} \emph {et~al.}}]{ATLAS:2022mgx}%
  \BibitemOpen
  \bibfield  {author} {\bibinfo {author} {\bibfnamefont {G.}~\bibnamefont {Aad}} \emph {et~al.} (\bibinfo {collaboration} {ATLAS}),\ }\bibfield  {title} {\bibinfo {title} {{Measurement of the total cross section and $\rho $-parameter from elastic scattering in pp collisions at $\sqrt{s}=13$~TeV with the ATLAS detector}},\ }\href {https://doi.org/10.1140/epjc/s10052-023-11436-8} {\bibfield  {journal} {\bibinfo  {journal} {Eur. Phys. J. C}\ }\textbf {\bibinfo {volume} {83}},\ \bibinfo {pages} {441} (\bibinfo {year} {2023})},\ \Eprint {https://arxiv.org/abs/2207.12246} {arXiv:2207.12246 [hep-ex]} \BibitemShut {NoStop}%
\bibitem [{\citenamefont {Verlinde}(2011)}]{Verlinde:2010hp}%
  \BibitemOpen
  \bibfield  {author} {\bibinfo {author} {\bibfnamefont {E.~P.}\ \bibnamefont {Verlinde}},\ }\bibfield  {title} {\bibinfo {title} {{On the Origin of Gravity and the Laws of Newton}},\ }\href {https://doi.org/10.1007/JHEP04(2011)029} {\bibfield  {journal} {\bibinfo  {journal} {JHEP}\ }\textbf {\bibinfo {volume} {04}},\ \bibinfo {pages} {029}},\ \Eprint {https://arxiv.org/abs/1001.0785} {arXiv:1001.0785 [hep-th]} \BibitemShut {NoStop}%
\bibitem [{\citenamefont {Fursaev}(2010)}]{Fursaev:2010ix}%
  \BibitemOpen
  \bibfield  {author} {\bibinfo {author} {\bibfnamefont {D.~V.}\ \bibnamefont {Fursaev}},\ }\bibfield  {title} {\bibinfo {title} {{`Thermodynamics' of Minimal Surfaces and Entropic Origin of Gravity}},\ }\href {https://doi.org/10.1103/PhysRevD.82.064013} {\bibfield  {journal} {\bibinfo  {journal} {Phys. Rev. D}\ }\textbf {\bibinfo {volume} {82}},\ \bibinfo {pages} {064013} (\bibinfo {year} {2010})},\ \bibinfo {note} {[Erratum: Phys.Rev.D 86, 049903 (2012)]},\ \Eprint {https://arxiv.org/abs/1006.2623} {arXiv:1006.2623 [hep-th]} \BibitemShut {NoStop}%
\bibitem [{\citenamefont {Bisognano}\ and\ \citenamefont {Wichmann}(1975)}]{Bisognano:1975ih}%
  \BibitemOpen
  \bibfield  {author} {\bibinfo {author} {\bibfnamefont {J.~J.}\ \bibnamefont {Bisognano}}\ and\ \bibinfo {author} {\bibfnamefont {E.~H.}\ \bibnamefont {Wichmann}},\ }\bibfield  {title} {\bibinfo {title} {{On the Duality Condition for a Hermitian Scalar Field}},\ }\href {https://doi.org/10.1063/1.522605} {\bibfield  {journal} {\bibinfo  {journal} {J. Math. Phys.}\ }\textbf {\bibinfo {volume} {16}},\ \bibinfo {pages} {985} (\bibinfo {year} {1975})}\BibitemShut {NoStop}%
\bibitem [{\citenamefont {Bisognano}\ and\ \citenamefont {Wichmann}(1976)}]{Bisognano:1976za}%
  \BibitemOpen
  \bibfield  {author} {\bibinfo {author} {\bibfnamefont {J.~J.}\ \bibnamefont {Bisognano}}\ and\ \bibinfo {author} {\bibfnamefont {E.~H.}\ \bibnamefont {Wichmann}},\ }\bibfield  {title} {\bibinfo {title} {{On the Duality Condition for Quantum Fields}},\ }\href {https://doi.org/10.1063/1.522898} {\bibfield  {journal} {\bibinfo  {journal} {J. Math. Phys.}\ }\textbf {\bibinfo {volume} {17}},\ \bibinfo {pages} {303} (\bibinfo {year} {1976})}\BibitemShut {NoStop}%
\end{thebibliography}%

\clearpage
\onecolumngrid
\section*{Supplemental Material}

\newcommand{\beginsupplement}{%
  \setcounter{table}{0}   \renewcommand{\thetable}{S\arabic{table}}%
  \setcounter{figure}{0}  \renewcommand{\thefigure}{S\arabic{figure}}%
  \setcounter{equation}{0}\renewcommand{\theequation}{S\arabic{equation}}}

\beginsupplement

\begin{figure}[htb]
  \centering
  \resizebox{0.28\columnwidth}{!}{%
    \begin{tikzpicture}[scale=1.25]
      \def\Rsmall{1.1}
      \def\Rbig  {2.4}

      \fill[blue!15,opacity=0.35] (0,0) circle (\Rsmall); 
      \draw[very thick] (0,0) circle (\Rsmall);           
      \draw[thick]      (0,0) circle (\Rbig);             

      \filldraw[black] (0,0) circle (0.6pt);

      \draw[->,thick] (0,0) -- (\Rsmall,0)
           node[midway,below=1pt] {$r$};

      \node at (-0.2,0.4) {$A$};
      \node at (1.55,0.55) {$B$};
      \node at (0,\Rsmall+0.25) {$\Sigma_{\perp}$};
    \end{tikzpicture}}
  \caption{Region \(A\) is bounded by the entangling surface
           \(\Sigma_{\perp}\) (the inner circle is a 1D illustration of the 2D surface); its complement is region \(B\).
           The arrow denotes the radial coordinate $r$ from the center of
           \(A\) to \(\Sigma_{\perp}\) localized at $r=R=R_{\!EE}$.}
  \label{fig:A-B-partition}
\end{figure}

\emph{Derivation of Eq.~\eqref{eq:Weyl_intro}.—}
We work on the \(t=0\) hypersurface and write
\(x^{\mu}=(t,\rp,\rpp)\) with
\(\rp\equiv x^{1}\in\mathbb{R}\) the longitudinal coordinate and
\(\rpp\equiv(x^{2},x^{3})\in\mathbb{R}^{2}\) the transverse vector
(\(\bm x\equiv(\rp,\rpp)\)).
Its magnitude is \(\rppmag=|\rpp|=b_\perp\) which can be identified with the impact-parameter space, and the three-dimensional radius is
\(r=\sqrt{\rp^{2}+\rppmag^{2}}\).
The volume element factorizes into longitudinal and transverse parts,
\[
\diff^{3}r=\diff\rp\,\diff^{2}r_{\perp},
\qquad
\diff^{2}r_{\perp}=\rppmag\,\diff\rppmag\,\diff\phi ,
\]
while in spherical coordinates \((r,\theta,\phi)\)
\(\diff^{3}r=r^{2}\sin\theta\,\diff r\,\diff\theta\,\diff\phi\).

The flat metric in longitudinal–transverse coordinates is
\begin{equation}
  \diff s^{2}
  = -\,\diff t^{2}
    + \diff\rp^{2}
    + \diff\rppmag^{2}
    + \rppmag^{2}\,\diff\phi^{2},
  \label{eq:metric_cart}
\end{equation}
and, with the substitutions
\(\rppmag = r\sin\theta\) and \(\rp = r\cos\theta\), this becomes the familiar
spherical-coordinate form
\begin{equation}
  \diff s^{2}
  = -\,\diff t^{2}
    + \diff r^{2}
    + r^{2}\bigl(\diff\theta^{2}
                + \sin^{2}\theta\,\diff\phi^{2}\bigr).
  \label{eq:metric_sph}
\end{equation}

In a neighborhood of the entangling surface \(\Sigma_{\perp}\), shown in Fig.~\ref{fig:A-B-partition}, an
arbitrary smooth metric admits the expansion
\begin{equation}\label{eq:metric_expansion}
ds^{2}=g_{\mu\nu}dx^{\mu}dx^{\nu}
      =-dt^2 + dr^{2}+\bigl[h_{ab}(x_{\perp})+\mathcal{O}(r^{2})\bigr]
       dx_{\perp}^{a}dx_{\perp}^{b},
\end{equation}
where \(h_{ab}\) is the induced metric on
\(\Sigma_{\perp}\). The
\(\mathcal{O}(r^{2})\) terms encode curvature corrections away from the
surface. Throughout, \(T(x)\equiv\langle T^{\mu}{}_{\mu}(x)\rangle\).
Under a constant Weyl rescaling \(g_{\mu\nu}\!\to\!e^{2\sigma}g_{\mu\nu}\)
the entangling sphere radius rescales as \(R\to e^{\sigma}R\).

\smallskip\noindent
\textbf{1. Modular Hamiltonian after tracing out the exterior.}
Tracing over all quark- and gluon-field modes with support \emph{outside}
a sphere of radius \(R\), as in Fig.~\ref{fig:A-B-partition}, produces the reduced state
\(\hat\rho_{<}(R)=\mathrm{Tr}_{r>R}|0\rangle\!\langle0|\).
Its generator of modular flow, \(K(R)\equiv-\ln\hat\rho_{<}(R)\),
governs time evolution for an observer confined to \(r<R\).
For the planar entangling surface that bounds the right Rindler wedge
\(\mathcal R^{+}=\{x^{1}>|t|\}\)\,\cite{Kabat:1994vj},
the Bisognano–Wichmann theorem
\cite{Bisognano:1975ih,Bisognano:1976za} identifies the modular
Hamiltonian with the boost generator,
\[
K_{\mathcal R^{+}}
  = 2\pi\!\int\!\diff^{3}x\,\bigl[x^{1}T^{00}-tT^{01}\bigr].
\]
Setting \(t=0\) one finds
\begin{equation}
K_{\mathcal R^{+}}
 = 2\pi\!\int_{x^{1}>0}\!\diff\rp\,\diff^{2}r_{\perp}\;
      \rp\,T_{00}(0;\rp,\rpp).
\label{eq:RindlerK}
\end{equation}
Because \(K\) couples linearly to \(T_{00}\), the eigenvalue density of
\(\hat\rho_{<}\) (the \emph{entanglement spectrum}) is fixed by
\(\langle T_{00}\rangle\). In the Lorentz-invariant vacuum with the
mostly-plus Minkowski metric,
\(T_{00}=-\tfrac{1}{4}T^{\mu}{}_{\mu}\); tracing Lorentz
indices therefore ties the spectrum directly to the trace spectral
density \(\rho_h\).

\smallskip\noindent
\textbf{2. Conformal map to a ball.}
A special conformal transformation with
\(a^{\mu}=(0,0,0,1/R)\) maps the half-space \(\rp>0\) to the ball
\(r<R\) and rescales the spherical measure \(r^{2}\sin\theta\):
\[
\bigl|\partial(\rp,\rpp)/\partial(r,\theta,\phi)\bigr|
      =\Bigl(\tfrac{2R}{R^{2}+r^{2}}\Bigr)^{4}.
\]
Because \(T_{00}\) has conformal weight 4, this Jacobian cancels and one
obtains the \emph{sphere-local} modular Hamiltonian
\cite{Casini:2011kv,Rosenhaus:2014woa}
\begin{equation}
K(R)
 = 2\pi
   \!\!\int_{r<R}
   \!\!\diff\rp\,\diff^{2}r_{\perp}\;
   \frac{R^{2}-r^{2}}{2R}\;
   T_{00}(0;\rp,\rpp).
\label{eq:localK}
\end{equation}
Evaluating its vacuum expectation value and tracing indices projects
\(T_{00}\) onto \(T^{\mu}{}_{\mu}\), reproducing the spectral density
\(\rho_{h}(r_\perp)\) that drives the mechanical entanglement entropy.

\smallskip\noindent
\textbf{3. Contact term.}
Energy–momentum conservation,
\(\partial_\mu T^{\mu\nu}=0\), implies the Ward identity
\[
\partial_\mu\bigl\langle T^{\mu\nu}(x)\,T_{00}(y)\bigr\rangle
  = -\delta^{(3)}(\bm x-\bm y)\,
    \partial_y^{\nu}\!\bigl\langle T_{00}(y)\bigr\rangle .
\]
Integrating the \(\nu=0\) component over \(t\) at equal times isolates a
\emph{purely local} distributional contribution. Tracing Lorentz
indices on the first operator fixes the unique contact piece of the
correlator \cite{Ben-Ami:2015zsa},
\begin{align}
\langle T(0,\bm x)\,T_{00}(0;\rp',\rpp')\rangle
  &= \langle T(0,\bm x)\rangle\,
     \delta(\rp-\rp')\,\delta^{(2)}(\rpp-\rpp')
     \,\,+\,\,\text{regular terms for }x\neq y.
\label{eq:Contact}
\end{align}

\smallskip\noindent
\textbf{4. Surface-local two-point function.}
Inserting \eqref{eq:Contact} into \eqref{eq:localK} gives
\begin{align}
\langle T(0,\bm x^{\!*})\,K\rangle
 &= 2\pi
    \!\!\int_{r'^{2}<R^{2}}
    \!\!\!\!\!\diff\rp'\,\diff^{2}r'_{\perp}\;
      \frac{R^{2}-\rp'^{2}-r_{\perp}^{\prime 2}}{2R}\,
      T(0,\bm x^{\!*})\,
      \delta(\rp^{\!*}\!-\rp')\,\delta^{(2)}(\rpp^{\!*}\!-\rpp') \notag \\[4pt]
 &= 2\pi\,T(0,\bm x)\,F(r)\bigl|_{r=r^{\!*}}, \label{eq:V1}
\end{align}
where \(F(r)=w(r)\,\Theta(R-r)\) with
\(w(r)=\tfrac{1}{2R}(R^{2}-r^{2})\).

For any spherically symmetric kernel \(F(r)\),
\[
\nabla^{2}F(r)
 = \frac{1}{r^{2}}\frac{\partial}{\partial r}
   \!\Bigl(r^{2}\,\frac{\partial F}{\partial r}\Bigr),
\]
so \(-\tfrac{4\pi R^{2}}{2}\nabla^{2}\) \emph{projects} the weight function onto the
surface term dictated by Gauss’ theorem, isolating the local contact
piece on the entangling sphere \(r=R\).
Differentiating yields
\be
-\tfrac{4\pi R^{2}}{2}\nabla^{2}F(r)=
  \underbrace{\frac{3\times 4\pi R^{2}}{2R}\,\Theta(R-r)}_{\text{bulk term}}
  +\underbrace{\delta(R-r)}_{\delta_{\Sigma}(r)}.\label{tdf}
\ee
The bulk contribution integrates to a power divergence (e.g.\ area-law \(\propto R^2/\epsilon^2\), curvature terms on \(\Sigma_\perp\), etc.) removable by counter-terms and therefore does not affect the universal logarithmic coefficient. The coefficient of $\delta(R-r)$ is unity by construction (Gauss’ theorem).
Using \eqref{tdf} in \eqref{eq:V1} and keeping only the surface term,
\begin{equation}
\langle T(0,\bm x)\,K\rangle
  = 2\pi\,T(0,\bm x)\,\delta(R-r).
\label{eq:SurfaceCorr}
\end{equation}
The factor $\delta(R-r)\equiv\delta_{\Sigma}(r)$ is a \emph{surface delta function} normalized to $\int\!\diff^{3}r\,\delta_{\Sigma}(r)=4\pi R^2$: it collapses any 3D integral onto the 2D entangling sphere $r=R$. Here $\delta_\Sigma^{(3D)}(r)=\delta(R-r)$ so $\int d^3r\,\delta_\Sigma^{(3D)}=4\pi R^2$, while in the main text~\eqref{eq:occupational} we use $\delta_\Sigma^{(2D)}(b_\perp)=\delta(b_\perp-R)/(2\pi R)$ so $\int d^2b_\perp\,\delta_\Sigma^{(2D)}=1$.
Concretely,
\begin{equation}
\int\!\diff^{3}r\,\delta_{\Sigma}(r)\,f(r,\Omega)
      = R^{2}\!\int_{r=R}\!\diff\Omega\,f(R,\Omega)
      = \int_{\Sigma_\perp}\!\sqrt{h}\,f,
\end{equation}
so only the values of the integrand on $\Sigma_{\perp}$ contribute.

Although Eq.~\eqref{eq:SurfaceCorr} was established within a conformal field theory, the derivation relies solely on locality and the Ward identity $\partial_{\mu}T^{\mu\nu}=0$; it therefore applies to any relativistic quantum field theory, including QCD.
We thus adopt \eqref{eq:SurfaceCorr} as the starting point for the explicit calculation of the entanglement–entropy flow that culminates in Eq.~\eqref{eq:FlowApp}.

\smallskip\noindent
\textbf{5. Entanglement-entropy flow.}
A uniform Weyl shift \(\delta g_{\mu\nu}=2\sigma g_{\mu\nu}\) gives
\cite{Rosenhaus:2014nha}
\be
\delta S_{\!EE}
  = -\sigma
    \!\int\!\diff^{3}x\,\langle T(0,\bm x)\,K\rangle
  = -\sigma\,2\pi
    \!\int_{r=R}\!\diff^{2}r_{\perp}\;
      T(0;\rp,\rpp),
\label{eq:FlowApp}
\ee
where we used \eqref{eq:SurfaceCorr} to collapse the 3D integral to 2D. Choosing \(\sigma=\delta R/R\) gives the RG-flow equation
\begin{equation}
R\,\frac{\diff S_{\!EE}}{\diff R}
  = -\,2\pi
    \!\int_{r=R}\!\diff^{2}r_{\perp}\;
      T\bigl(0;\rp,\rpp\bigr).
\label{eq:SEEflow}
\end{equation}
Integrating from \(\sUV\) to \(R\), and discarding the analytic (counter-term) piece yields the universal logarithm:
\begin{equation}
\boxed{S_{\!EE}^{\text{univ}}(R)
 = \,2\pi\,
     \ln\!\frac{R}{\sUV}
     \int_{r=R}\!\diff^{2}r_{\perp}\;
        T\bigl(0;\rp,\rpp\bigr)=\,2\pi\,
     \ln\!\frac{R}{\sUV}
     \int_{r=R}\!R^2\,\diff\Omega\;
        T\bigl(0;R,\Omega\bigr)=\,2\pi\,
     \ln\!\frac{R}{\sUV}
     \int_{\Sigma_\perp}R^2\,\!\diff\theta\diff\phi\sqrt{h}\,
        T\bigl(0;R,\Omega\bigr),\,}
\label{eq:SunivFinal}
\end{equation}
with \(T(x)\equiv\langle T^{\mu}{}_{\mu}(x)\rangle\), and using the spherical coordinate \eqref{eq:metric_sph} in the last two equations. Eq.~\eqref{eq:SunivFinal} reproduces Eq.~\eqref{eq:Weyl_intro} of the main text. Note that at $r=R$, the induced metric $h_{ab}$ on the two-sphere (using the spherical coordinate \eqref{eq:metric_sph}) is given by:
\begin{equation}
 \left.\diff s^{2}\right|_{t=0,r=R}=R^2h_{ab}dx^{a}dx^{b}=R^{2}\bigl(\diff\theta^{2}
                + \sin^{2}\theta\,\diff\phi^{2}\bigr)\,.
  \label{eq:metric_sph2}
\end{equation}
with its determinant $h=\sin^{2}\theta$.

\bigskip
\emph{Mechanical core radius and scalar/energy rms radii from lattice GFF.—}
We work with the sea-quark and gluon contributions to the proton EMT,
\(\,T_{q,g}^{\mu\nu}\), whose matrix elements are parameterized by
the gravitational form factors (GFFs) as in Eq.~\eqref{eq:EMT_nucl},
with \(t=(P'-P)^2=-K^2\).
Using the scalar–tensor separation of GFFs, the scalar ($0^{++}$) combination is
\be
A^{S}_{q,g}(t)\equiv
A_{q,g}(t)-\frac{3t}{4m_N^{2}}\,D_{q,g}(t)\,,
\ee
while \(A_{q,g}(t)\) carries the traceless ($2^{++}$) part.
Throughout this section, \emph{all} radii quoted and plotted in
Fig.~\ref{fig:radii-comparison} of the main text are extracted
\emph{separately for quarks and for gluons}; no \(q\!+\!g\) sum is taken
when computing the individual radii.

\subsubsection*{Mechanical core radius from the Breit-frame pressure node}
\begin{figure}[t]
  \centering
  \includegraphics[width=0.4\textwidth]{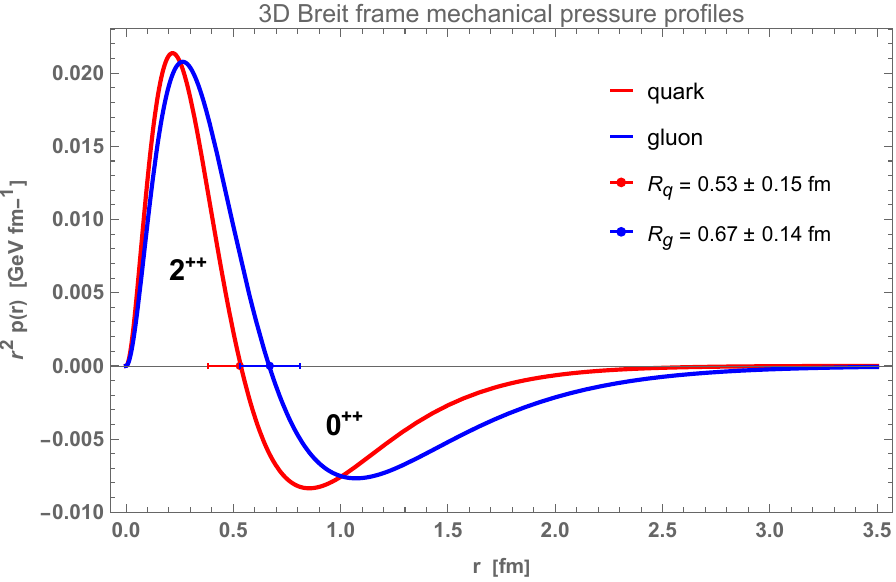}
  \caption{Radial distribution \(r^{2}p_{q,g}(r)\) normalized to
  \(A_{q,g}(0)\), computed from Eq.~\eqref{eq:pressureAnalytic}.
  The node marks the mechanical equilibrium radius \(R_{q,g}\) where
  repulsive ($2^{++}$) and attractive ($0^{++}$) stresses balance. \label{fig:pressure}}
\end{figure}

The isotropic pressure density in the Breit frame is
\begin{equation}
p_{q,g}(r)
=\frac{1}{6m_N\,r^2}\frac{d}{dr}
 \!\left[r^2\frac{d}{dr}D_{q,g}(r)\right],
\label{eq:pressure-def}
\end{equation}
and obeys the stability (von Laue) condition
\(\int_0^\infty\!dr\,4\pi r^2\,p_{q,g}(r)=0\).
The \emph{mechanical core radius} \(R_{q,g}\) is defined by the node of
the surface force \(4\pi r^2 p_{q,g}(r)\),
i.e.\ \(p_{q,g}(R_{q,g})=0\), where the repulsive ($2^{++}$) and
attractive ($0^{++}$) stresses balance, see Fig.~\ref{fig:pressure}.

For the two–scale dipole representation used in Table~\ref{tab:gff_params},
\[
A_{q,g}(t)=\frac{\alpha_{A,i}}{(1-t/\Lambda_{A,q/g}^{2})^{2}},\qquad
A^{S}_{q,g}(t)=\frac{\alpha_{S,i}}{(1-t/\Lambda_{S,q/g}^{2})^{2}},
\]
(with \(\alpha_{S,i}\!\simeq\!\alpha_{A,i}\) in the lattice fits),
the corresponding pressure admits a simple analytic form,\footnote{This closed form follows from the standard 3D Fourier
transform of a dipole, together with
\(D_{q,g}(t)=\tfrac{4m_N^2}{3t}\!\left[A_{q,g}(t)-A^{S}_{q,g}(t)\right]\).}
\begin{equation}
p_{q,g}(r)=\frac{\alpha_{A,i}\,m_N}{36\pi}
\Bigl[\Lambda_{A,q/g}^{3}\,e^{-\Lambda_{A,q/g} r}
     -\Lambda_{S,q/g}^{3}\,e^{-\Lambda_{S,q/g} r}\Bigr],
\label{eq:pressureAnalytic}
\end{equation}
which satisfies the von Laue condition identically.
The node condition \(p_{q,g}(R_{q,g})=0\) gives the mechanical core
radius in closed form,
\begin{equation}
R_{q,g}=\frac{3\,\ln(\Lambda_{A,q/g}/\Lambda_{S,q/g})}
              {\Lambda_{A,q/g}-\Lambda_{S,q/g}}\;,
\quad
\text{(in } \mathrm{GeV}^{-1}\text{; multiply by } \hbar c=0.19733~\text{fm/GeV)}.
\label{eq:Rcore-analytic}
\end{equation}
The values used in Fig.~\ref{fig:radii-comparison} are obtained by
inserting the dipole masses \((\Lambda_{A,q/g},\Lambda_{S,q/g})\) from
Table~\ref{tab:gff_params}.

\subsubsection*{Scalar and energy rms radii (quark/gluon \emph{separately})}
\paragraph*{Definitions.}
The slope of the Lorentz-invariant \emph{scalar} form factor
\(A^{S}_{q,g}(t)\equiv\eta_{\mu\nu}T^{\mu\nu}_{q,g}(t)\)
at the origin defines a frame-independent rms radius, while the slope of
the \emph{energy density}
\(A^{00}_{q,g}(t)=T^{00}_{q,g}(t)\)—taken in the nucleon rest
frame—defines a distinct Breit-frame radius:
\begin{align}
\bigl\langle r^{2}_{s,q(g)}\bigr\rangle &=
-6\,
 \left.\frac{dA^{S}_{q,g}(t)}{dK^{2}}\right|_{K^{2}=0},
\quad
\text{with }t=-K^{2},
\nonumber\\[-1pt]
\bigl\langle r^{2}_{00,q(g)}\bigr\rangle &=
-6\,
 \left.\frac{dA^{00}_{q,g}(t)}{dK^{2}}\right|_{K^{2}=0},
\label{eq:radii-def}
\end{align}
where
\[
\begin{aligned}
A^{S}_{q,g}(t) &=
A_{q,g}(t)
 -\frac{3t}{4m_N^{2}}\,
  D_{q,g}(t),
\\[2pt]
A^{00}_{q,g}(t) &=
A_{q,g}(t)
 -\frac{t}{4m_N^{2}}\,
  D_{q,g}(t).
\end{aligned}
\]

\paragraph*{Two–scale dipole result.}
Within the two–scale dipole model used in
Table~\ref{tab:gff_params},\;\;
\(A^{S}_{q,g}(t)=
\alpha_{S,i}/(1-t/\Lambda_{S,q/g}^{2})^{2}\),
\(
A_{q,g}(t)=\alpha_{A,i}/(1-t/\Lambda_{A,q/g}^{2})^{2}
\),
and \(D_{q,g}(t)=\tfrac{4m_N^{2}}{3t}\!\left[A_{q,g}(t)-A^{S}_{q,g}(t)\right]\).
For \(\alpha_{S,i}=\alpha_{A,i}\) (as in current lattice fits) one finds the compact analytic expressions
\begin{equation}
\boxed{\
\langle r^{2}_{s,q(g)}\rangle
      =\frac{12}{\Lambda_{S,q/g}^{2}},\qquad
\langle r^{2}_{00,q(g)}\rangle
      =\frac{8}{\Lambda_{A,q/g}^{2}}
       +\frac{4}{\Lambda_{S,q/g}^{2}}\, }
\label{eq:radii-analytic}
\end{equation}
in \(\mathrm{GeV}^{-2}\); multiply by
\(\hbar^{2}c^{2}=(0.19733~\mathrm{fm})^{2}\) to convert to \(\mathrm{fm}^{2}\).
These formulas are applied \emph{separately} to the quark and to the
gluon sectors using \(\Lambda_{A,q/g},\Lambda_{S,q/g}\) from
Table~\ref{tab:gff_params}, yielding the individual rms radii displayed in Fig.~\ref{fig:radii-comparison}.

\bigskip
\emph{Lattice GFF fit parameters.—}%
Here we summarize the lattice fit parameters for the gravitational form factors (GFF) used in the main text.
\FloatBarrier
\begin{table}[!h]
\caption{Dipole [$A_X(t)$] and $D$-term [$D_X(t)$] fit parameters from
lattice-QCD analyses~\cite{Hackett:2023rif,Hackett:2023nkr}.
The first block lists nucleon quark (\(q\)) and gluon (\(g\)) sectors;
the second block shows the corresponding pion sectors.
Masses are in GeV and uncertainties are statistical.
Scalar form factors are constructed as
\(A_N^{S}=A_q^{S}+A_g^{S}\) for the nucleon and
\(A_\pi^{S}=A_q^{S}+A_g^{S}\) at \(m_\pi=170~\text{MeV}\).
For completeness, the nucleon scalar-pole masses (that reproduce the lattice D-term) are
\(\Lambda_{S,q}=0.804(108)\,\text{GeV}\) and
\(\Lambda_{S,g}=0.596(77)\,\text{GeV}\).}
\label{tab:gff_params}
\scriptsize
\setlength{\tabcolsep}{3.9pt}
\begin{ruledtabular}
\begin{tabular}{lcccccc}
Species & Sector & $\alpha_A=\alpha_S$ & $\Lambda_A$ &
                         $\alpha_D$ & $\Lambda_D$ \\ \hline
\multirow{2}{*}{Nucleon} & $q$ & 0.510(25) & 1.477(44) & $-1.30(49)$ & 0.81(14) \\
                         & $g$ & 0.501(27) & 1.262(18) & $-2.57(84)$ & 0.538(65) \\[1pt] \hline
\multirow{2}{*}{Pion}    & $q$ & 0.481(15) & 1.262(37) & $-0.304(26)$ & 1.44(21) \\
                         & $g$ & 0.546(18) & 1.129(41) & $-0.596(65)$ & 0.677(65) \\
\end{tabular}
\end{ruledtabular}
\end{table}

\end{document}